\pdfoutput=1

\documentclass[a4paper,cits]{JINST}
\usepackage{lineno}
\usepackage{microtype}
\usepackage{booktabs}
\usepackage{amsmath}
\usepackage[hyphens]{url}

\title{Ionization and scintillation response of high-pressure xenon gas to alpha particles}

\author{
The NEXT Collaboration

V.~\'Alvarez,$^{a}$
F.I.G.~Borges,$^{b}$
S.~C\'arcel,$^{a}$
S.~Cebri\'an,$^{c}$
A.~Cervera,$^{a}$
C.A.N.~Conde,$^{b}$
T.~Dafni,$^{c}$
J.~D\'iaz,$^{a}$
M.~Egorov,$^{d}$
R.~Esteve,$^{e}$
P.~Evtoukhovitch,$^{f}$
L.M.P.~Fernandes,$^{b}$
P.~Ferrario,$^{a}$
A.L.~Ferreira,$^{g}$
E.D.C.~Freitas,$^{b}$
V.M.~Gehman,$^{d}$
A.~Gil,$^{a}$
A.~Goldschmidt,$^{d}$
H.~G\'omez,$^{c}$
J.J.~G\'omez-Cadenas,$^{a}$
D. Gonz\'alez-D\'iaz,$^{c}$
R.M.~Guti\'errez,$^{h}$
J.~Hauptman,$^{i}$
J.A.~Hernando Morata,$^{j}$
D.C.~Herrera,$^{c}$
I.G.~Irastorza,$^{c}$
M.A.~Jinete,$^{h}$
L.~Labarga,$^{k}$ 
A.~Laing,$^{a}$
I.~Liubarsky,$^{a}$
J.A.M.~Lopes,$^{b}$
D.~Lorca,$^{a}$
M.~Losada,$^{h}$
G.~Luz\'on,$^{c}$
A.~Mar\'i,$^{e}$
J.~Mart\'in-Albo,$^{a}$
T.~Miller,$^{d}$
A.~Moiseenko,$^{f}$
F.~Monrabal,$^{a}$
C.M.B.~Monteiro,$^{b}$
F.J.~Mora,$^{e}$
L.M. Moutinho,$^{g}$
J.~Mu\~noz~Vidal,$^{a}$
H.~Natal da Luz,$^{b}$
G.~Navarro,$^{h}$
M.~Nebot-Guinot,$^{a}$
D.~Nygren,$^{d}$
C.A.B.~Oliveira,$^{d}$
R.~Palma,$^{l}$
J.~P\'erez,$^{k}$
J.L.~P\'erez~Aparicio,$^{l}$
J.~Renner,$^{d}$
L.~Ripoll,$^{m}$
A.~Rodr\'iguez,$^{c}$
J.~Rodr\'iguez,$^{a}$
F.P.~Santos,$^{b}$
J.M.F.~dos~Santos,$^{b}$
L.~Segu\'i,$^{c}$
L.~Serra,$^{a}$
D.~Shuman,$^{d}$
A.~Sim\'on,$^{a}$
C.~Sofka,$^{n}$
M.~Sorel,$^{a}$\thanks{Corresponding author}~
J.F.~Toledo,$^{d}$
A.~Tom\'as,$^{c}$
J.~Torrent,$^{m}$
Z.~Tsamalaidze,$^{f}$
D.~V\'azquez,$^{j}$
J.F.C.A.~Veloso,$^{g}$
R.~Webb,$^{n}$
J.T.~White,$^{n}$
N.~Yahlali$^{a}$\\
\llap{$^{a}$}
Instituto de F\'isica Corpuscular (IFIC), CSIC \& Universitat de Val\`encia\\
Calle Catedr\'atico Jos\'e Beltr\'an, 2, 46980 Paterna, Valencia, Spain\\
\llap{$^{b}$}
Departamento de Fisica, Universidade de Coimbra\\
Rua Larga, 3004-516 Coimbra, Portugal\\
\llap{$^c$}
Lab.\ de F\'isica Nuclear y Astropart\'iculas, Universidad de Zaragoza\\ 
Calle Pedro Cerbuna, 12, 50009 Zaragoza, Spain\\
\llap{$^d$}
Lawrence Berkeley National Laboratory (LBNL)\\
1 Cyclotron Road, Berkeley, California 94720, USA\\
\llap{$^{e}$}
Instituto de Instrumentaci\'on para Imagen Molecular (I3M), Universitat Polit\`ecnica de Val\`encia\\ 
Camino de Vera, s/n, Edificio 8B, 46022 Valencia, Spain\\
\llap{$^{f}$}
Joint Institute for Nuclear Research (JINR)\\
Joliot-Curie 6, 141980 Dubna, Russia\\
\llap{$^{g}$}Institute of Nanostructures, Nanomodelling and Nanofabrication (i3N), Universidade de Aveiro\\
Campus de Santiago, 3810-193 Aveiro, Portugal\\
\llap{$^{h}$}
Centro de Investigaciones, Universidad Antonio Nari\~no\\ 
Carretera 3 este No.\ 47A-15, Bogot\'a, Colombia\\
\llap{$^{i}$}
Department of Physics and Astronomy, Iowa State University\\
12 Physics Hall, Ames, Iowa 50011-3160, USA\\
\llap{$^{j}$}
Instituto Gallego de F\'isica de Altas Energ\'ias (IGFAE), Univ.\ de Santiago de Compostela\\
Campus sur, R\'ua Xos\'e Mar\'ia Su\'arez N\'u\~nez, s/n, 15782 Santiago de Compostela, Spain\\
\llap{$^{k}$}
Departamento de F\'isica Te\'orica, Universidad Aut\'onoma de Madrid\\
Ciudad Universitaria de Cantoblanco, 28049 Madrid, Spain\\
\llap{$^{l}$}
Dpto.\ de Mec\'anica de Medios Continuos y Teor\'ia de Estructuras, Univ.\ Polit\`ecnica de Val\`encia\\
Camino de Vera, s/n, 46071 Valencia, Spain\\
\llap{$^{m}$}
Escola Polit\`ecnica Superior, Universitat de Girona\\
Av.~Montilivi, s/n, 17071 Girona, Spain\\
\llap{$^{n}$}
Department of Physics and Astronomy, Texas A\&M University\\
College Station, Texas 77843-4242, USA\\

E-mail: \email{sorel@ific.uv.es}
}

\abstract{
High-pressure xenon gas is an attractive detection medium for a variety of applications in fundamental and applied physics. In this paper we study the ionization and scintillation detection properties of xenon gas at 10 bar pressure. For this purpose, we use a source of alpha particles in the NEXT-DEMO time projection chamber, the large scale prototype of the NEXT-100 neutrinoless double beta decay experiment, in three different drift electric field configurations. We measure the ionization electron drift velocity and longitudinal diffusion, and compare our results to expectations based on available electron scattering cross sections on pure xenon. In addition, two types of measurements addressing the connection between the ionization and scintillation yields are performed. On the one hand we observe, for the first time in xenon gas, large event-by-event correlated fluctuations between the ionization and scintillation signals, similar to that already observed in liquid xenon. On the other hand, we study the field dependence of the average scintillation and ionization yields. Both types of measurements may shed light on the mechanism of electron-ion recombination in xenon gas for highly-ionizing particles. Finally, by comparing the response of alpha particles and electrons in NEXT-DEMO, we find no evidence for quenching of the primary scintillation light produced by alpha particles in the xenon gas.
}

\keywords{Gaseous detectors; Time projection Chambers (TPC); Charge transport and multiplication in gas; Scintillators, scintillation and light emission processes (solid, gas and liquid scintillators)} 

\begin{document}


\tableofcontents

\section{Introduction} \label{sec:Introduction}

\begin{figure}[t!b!]
\begin{center}
\includegraphics[scale=.35]{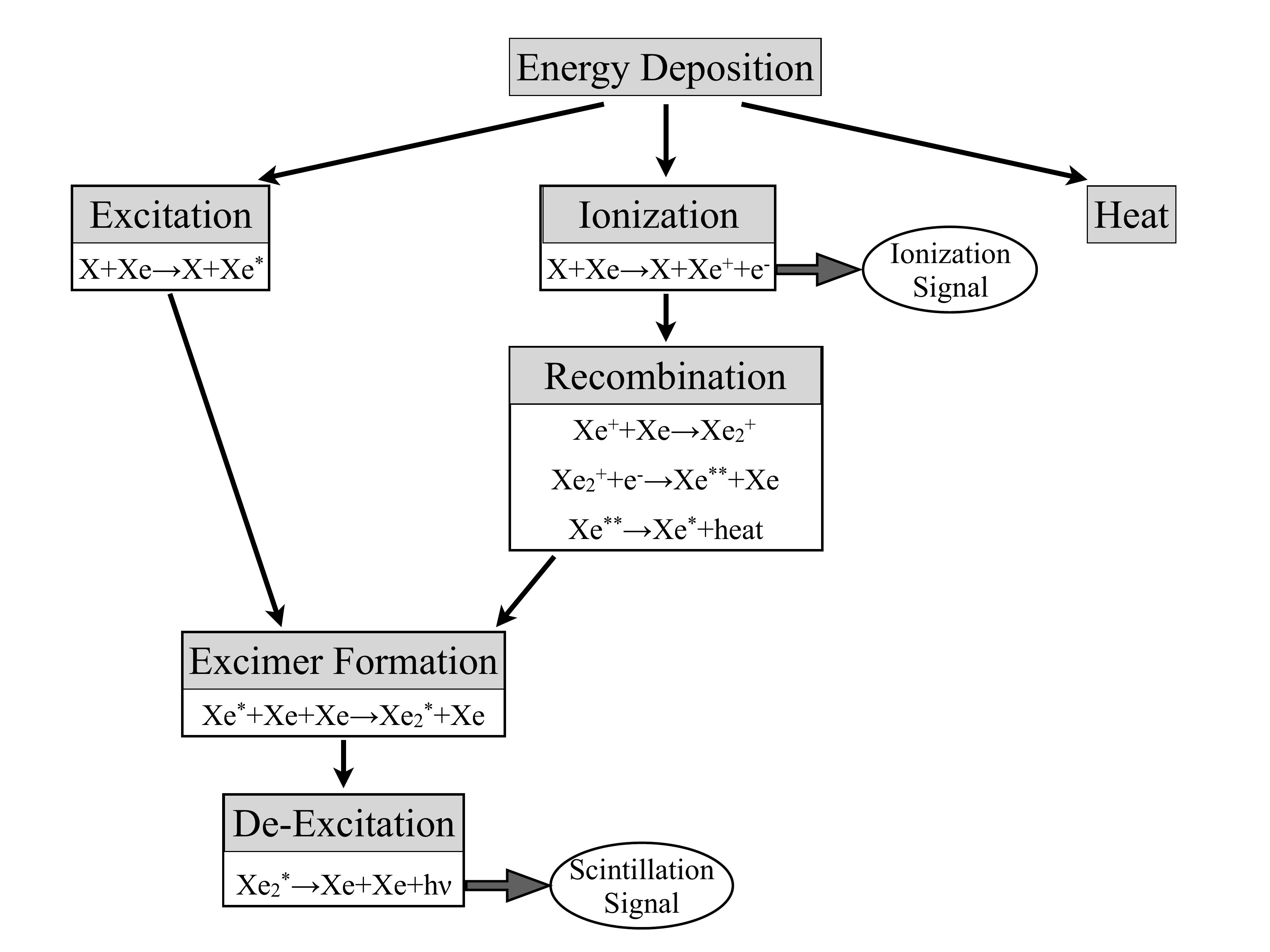}
\end{center}
\caption{Main processes responsible for the ionization and scintillation signals in xenon. The symbol X indicates ionizing radiation. Adapted from \cite{Aprile:2009dv}.}\label{fig:energyfate}
\end{figure}

Particles traversing a xenon-based detector transfer energy to xenon atoms in the form of heat, ionization and excitation, see Fig.~\ref{fig:energyfate}. The electron-ion pairs produced by the ionization process can give rise to a detectable signal, provided that a sufficiently strong electric field is applied to the detection medium to prevent electron-ion recombination. Scintillation photons emitted through the radiative decay of excited diatomic molecules (or excimers) of xenon can also be detected. The possibility to detect both the scintillation and ionization signals produced by the passage of particles is the main advantage of xenon as a detection medium, together with its relatively high stopping power. In addition, for xenon in the gas phase, the ionization fluctuations are characterized by a small intrinsic Fano factor \cite{Nygren:2007zzc}, permitting accurate calorimetry using the ionization signal alone. For these reasons, xenon-based detectors are used for a variety of applications in fundamental and applied physics, including searches for WIMP dark matter direct detection \cite{Akimov:2006qw,Abe:2013tc,Aprile:2011dd,Akerib:2012ys}, for neutrinoless double beta \cite{Auger:2012gs} and lepton-flavor-violating \cite{Adam:2009ci} decays, as well as X-ray astronomy \cite{Jahoda:2005uw}, gamma-ray astronomy \cite{Aprile:2008ft}, and medical imaging \cite{Giboni:2007zz,grignon}.

The scintillation (often called \textit{S1}) and ionization (\textit{S2}) signals measured by xenon-based detectors depend on the production rate of excited atoms Xe$^{\ast}$ and electron-ion pairs e$^-$-Xe$^+$ produced by ionizing radiation, as well as on the electron-ion recombination strength. As the recombination rate increases, the total scintillation yield becomes greater than the yield provided by direct excitation of xenon atoms (\textit{primary scintillation}), given that each recombined pair is accompanied by the emission of one scintillation photon (\textit{recombination luminescence}), see Fig.~\ref{fig:energyfate}. The scintillation and ionization signals also depend on the photon and ionization electron transport properties through the detector, respectively. The scintillation signal is also affected by the scintillation emission spectrum and time response. Because most xenon-based detectors make use of xenon in its liquid phase, the detection properties of liquid xenon \cite{Aprile:2009dv} are generally better understood than the ones for xenon gas \cite{AprileBook}. A brief summary of xenon gas detection properties follows. We restrict our summary to measurements performed with $\alpha$, $\beta$ or $\gamma$ radiation, where the total energy loss is dominated by electronic (as opposed to nuclear) energy loss. 

The scintillation and ionization yields are usually expressed in terms of the average energies necessary to create a primary scintillation photon ($W_{\mathrm s}$) and an electron-ion pair ($W_{\mathrm i}$), respectively. Typically, both values are taken as empirical parameters provided by experiments. The mean energy to form an electron-ion pair in high-pressure xenon, and when negligible recombination occurs, has been accurately measured to be about $W_{\mathrm i}=22$ eV using X-rays, gamma rays and electrons \cite{vinagre, platzman, hurst, Ahlen:1980xr}. The dependence of $W_{\mathrm i}$ on electric field and gas density has also been systematically studied. For gas densities lower than $\rho\sim 0.3$ g/cm$^3$ (gas pressures less than about 40 bars at room temperature, that is in the range of pressures of interest here), recombination effects are small and $W_{\mathrm i}$ does not depend on gas density \cite{Nygren:2009zz}. For very high gas densities, $\rho\gtrsim 0.3$ g/cm$^3$, $W_{\mathrm i}$ depends not only on the electric field but also on the gas density \cite{bolotnikov1997}. The understanding of the ionization yield in xenon gas for alpha particles is instead less complete. The much higher ionization density produced by alpha particles, together with more efficient recombination processes (see below), may suppress the ionization signal. The ionization quenching factor $q_{\mathrm i}$ for alpha particles is an empirical value defined as the ratio of the amount of ionization produced by an alpha particle to the amount of ionization produced by an electron (or gamma ray) for the same deposited energy. No ionization quenching ($q_{\mathrm i}(\alpha)=1$) is reported in \cite{jesse}, where the value of $W_{\mathrm i}=22.0$ eV for electrons is to be compared with essentially the same number ($W_{\mathrm i}=21.9$ eV) for alphas. These values were obtained with a ionization chamber operated in the saturation region, and filled with xenon gas at a pressure of about 1 bar. On the other hand, the authors of \cite{Luscher:1998sd} report an ionization quenching factor of $q_{\mathrm i}(\alpha)=0.15$ in a time projection chamber filled with xenon gas at 5 bar and for a $\sim$1 kV/cm drift field. The results in \cite{pushkin_methane} suggest that this large difference in $q_{\mathrm i}(\alpha)$ is most likely due to the presence of methane, in a 4\% proportion, in the xenon gas mixture used in \cite{Luscher:1998sd}.

The primary scintillation yield of xenon gas has traditionally been far less understood than the ionization yield. Recently, several measurements of $W_{\mathrm s}$ have become available, using X-rays and low-energy gamma rays. The value $W_{\mathrm s}=76\pm 12$ eV is reported in \cite{Parsons:1990hv} for a Xe(90\%)/He(10\%) mixture at 15 bar and for a drift field of order 0.1 kV/cm$\cdot$bar, with similar results for pure xenon. The authors of \cite{docarmo} obtain $W_{\mathrm s}=111\pm 16$ eV at 1 bar pressure and 0.35 kV/cm$\cdot$bar drift field. The value $W_{\mathrm s}=72\pm 6$ eV is extracted in \cite{Fernandes:2010gg}, for gas pressures in the 1--3 bar range and for drift fields in the 0.15--0.6 kV/cm$\cdot$bar range, with no significant variations with pressure and with field. Finally, the authors of \cite{Resnati:2012dk} measure $W_{\mathrm s}=26^{+7}_{-6}$ eV at 40 bar pressure and no electric field. The difference between these results is presently not fully understood, as it can be attributed only partially to the different gas density and drift field conditions. As for the ionization channel, the scintillation response to alpha particles may be quenched in xenon with respect to the maximum scintillation response for the same energy deposition. Scintillation light quenching is independent of electron-ion recombination effects which may also be present, and which would increase (rather than decrease) the scintillation light yield. The scintillation quenching factor $q_{\mathrm s}(\alpha)=0.77$ has been reported for alpha particles in liquid xenon \cite{dokelxescint,tanaka}. \textit{Bi-excitonic quenching} \cite{hitachi} has been proposed to explain the value $q_{\mathrm s}(\alpha)<1$ in liquid xenon. In this model, the concentration of excited atoms is sufficiently high for the usual excimer formation processs Xe$^{\ast}$+Xe$\to$Xe$_2^{\ast}$ in Fig.~\ref{fig:energyfate} to be partially replaced with Xe$^{\ast}$+Xe$^{\ast}\to$Xe$_2^{\ast\ast}\to$Xe+Xe$^+$+e$^-$. The latter process is more likely for xenon at high density and for radiation producing a high ionization density. This electron may then recombine to produce only one scintillation photon from two (rather than one) excited atoms, producing scintillation light quenching. Similar scintillation light quenching processes  may occur in high-density gaseous xenon as well, see for example \cite{kusano,bogdanov}. To our knowledge, however, no results on scintillation light quenching for alpha particles in gaseous xenon are available. 

Recombination in high-pressure xenon gas has been studied with alpha particles by measuring how the average scintillation and ionization yields vary as a function of drift field and gas density \cite{pushkin,bolotnikov1999, kobayashi,Saito:2003dz}. For example, the authors of \cite{kobayashi} and \cite{Saito:2003dz} measure that the number of scintillation photons decreases monotonously with the electric field, and that the scintillation yield in the field-independent region is about 40\% of that at zero electric field for $\gtrsim$10 bar xenon gas pressure. In addition, in liquid xenon, recombination effects have also shown to produce event-by-event, anti-correlated, fluctuations between the ionization and scintillation signals \cite{Conti:2003av, Aprile:2007qd}. For example, the authors of \cite{Aprile:2007qd} report correlation coefficients between the ionization and scintillation signals in the $-0.87$ to $-0.74$ range for 662 keV gamma rays, as the drift field is increased from 1 to 4 kV/cm. This fact has been exploited to obtain a better reconstruction of the deposited energy by combining the two observables. From the theory point of view, two approaches are typically used to try to understand recombination. The Jaffe model of recombination \cite{jaffe} assumes that the initial ionization charge is distributed in a column-shaped volume around the ionizing particle trajectory. Electrons and ions drift away from this column under the effects of drift field and diffusion. During the time evolution of the column, electrons can be captured by positive ions, in a process called \textit{volume} (or \textit{columnar}) recombination. The Onsager model of recombination \cite{Onsager:1938zz} is instead based on the concept of the so-called \textit{initial} (or \textit{geminate}) recombination of electron-ion pairs. It assumes that the electron has a finite probability of being captured back by the Coulomb field of its parent ion, and that this probability is reduced by the application of a drift electric field. According to \cite{bolotnikov1999}, a combination of both models is needed to describe the ionization and scintillation response of high-pressure xenon gas to alpha particles. 

The transport properties of ionization electrons as they drift in xenon gas, namely the electron drift velocity and diffusion, have also been systematically studied as a function of drift field and xenon gas pressure (see \cite{Peisert:1984xj} and references therein). In the presence of a drift field, electrons acquire a net motion in the direction of the field with a stationary drift velocity, $\vec{v}_d$. Diffusion is the spread of ionization electrons in the detector volume as the cloud of electrons is drifting. These macroscopic properties can be derived by solving the Boltzmann transport equations \cite{Fraser:1986tn}, provided that a good knowledge of the electron scattering differential cross sections as a function of electron energy is available. Different Monte Carlo toolkits for the simulation of electron drift and diffusion exist (see, for example, \cite{magboltz,Escada:2011cs}). To set the scale, for xenon gas at 10 bar pressure and for a 1 kV/cm drift field, electron drift velocities of about 1 mm/$\mu$s are obtained, together with a longitudinal diffusion spread of a few mm for a 1 m drift. 

No measurable effect on the scintillation signal can typically be attributed to photon transport properties in xenon gas. The self-absorption of scintillation photons in pure xenon gas is negligible, given that the emission spectrum due to excimer formation is well separated from the absorption spectrum of atomic xenon. For this reason, any evidence for scintillation light absorption in xenon-based detectors is typically attributed to molecular impurities in the gas.

This paper aims at improving our understanding of some of the xenon gas properties mentioned above. Measurements of ionization electron transport properties are described in Sec.~\ref{sec:ElectronTransport}, while electron-ion recombination studies are reported in Sec.~\ref{sec:Recombination}. A comparison of the primary scintillation light yield for alpha particles and electrons in xenon gas is also given in Sec.~\ref{sec:Recombination}. This work is performed in the context of the NEXT Collaboration R\&D, to build a 100 kg scale high-pressure xenon gas time projection chamber with the goal of performing a sensitive search neutrinoless double beta decay of the $^{136}$Xe isotope \cite{Granena:2009it,:2012haa}. Specifically, the data presented in this paper were obtained by exposing NEXT-DEMO, the large scale prototype for the NEXT-100 detector, to an alpha source. Before presenting our results (Sec.~\ref{sec:ElectronTransport} and \ref{sec:Recombination}), we describe the experimental setup in Sec.~\ref{sec:ExperimentalSetup}. The NEXT-DEMO data processing and selection of alpha candidate events is explained in Sec.~\ref{sec:ProcessingSelection}.

\section{Experimental setup} \label{sec:ExperimentalSetup}

\subsection{NEXT-DEMO detector} \label{subsec:ExperimentalSetupNextDemo}

\begin{figure}[t!b!]
\begin{center}
\includegraphics[scale=.30]{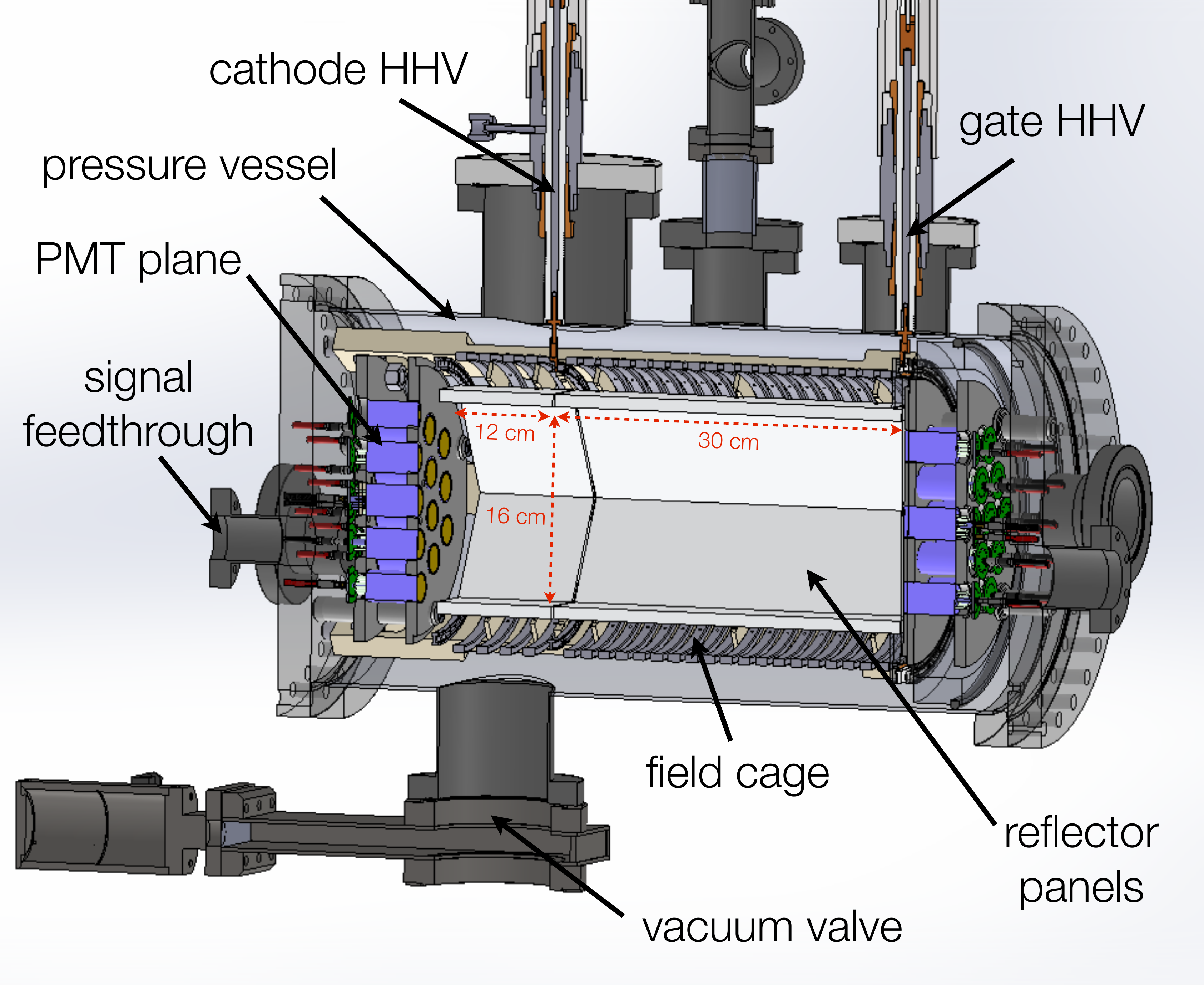}
\end{center}
\caption{Drawing of the NEXT-DEMO time projection chamber.}\label{fig:nextdemo}
\end{figure} 

NEXT-DEMO \cite{nextdemopaper} is a time projection chamber (TPC) filled with xenon gas at high pressure. A drawing of the detector is shown in Fig.~\ref{fig:nextdemo}. The TPC fits within a stainless steel pressure vessel, 600 mm long and 300 mm inner diameter, certified for up to 15 bar operational pressure as well as for operation under vacuum. The detector materials release electronegative impurities into the xenon gas, which need to be removed to avoid degradation of the detector performance. This is accomplished by continuously re-circulating the xenon gas through a getter at a flow of about 10 liters per minute. The TPC fiducial volume consists of a hexagonal prism. The apothem of the hexagon base is 80 mm. The six prism sides are covered with PTFE reflector panels. The fiducial volume extent along the third (drift) dimension, 300 mm long, is defined by metallic planes establishing the drift electric field. On one end, the cathode grid is made of 20 stainless steel wires at a 10 mm pitch. On the other end, the gate mesh defining the anode is made of crossed stainless steel wires at a 0.25 mm pitch. The maximum designed drift field is $\sim 1$~kV/cm. A region of \textit{electroluminescent} (EL, in the following) amplification of the ionization charge is defined by creating a high field region between the gate mesh and a second mesh at ground. The grounded mesh is made of crossed stainless steel wires at a 0.50 mm pitch. The passage of ionization electrons in the EL region excite xenon atoms, and a large number of VUV photons per ionization electron is produced as a result of excimer de-excitation. The ground mesh is placed at a 4.7 mm distance from the gate mesh. The maximum designed EL field between the two meshes is $\sim 4$~kV/(cm$\cdot$bar). The readout is made of a photomultiplier tube (PMT) plane, consisting of 19 Hamamatsu R7378A PMTs, located 120 mm behind the cathode grid. The light collection in this 120 mm long buffer region is increased via reflector panels similar to the drift region ones, but with no TPB coating on them. The PMTs, with a 21 mm diameter active area each, are placed in this cathode readout plane following the hexagonal pattern of the reflector panels, 35 mm apart from each other. The active area coverage within the hexagon is 40\%. These PMTs were chosen because they are capable of withstanding 20 bar pressure, and are sensitive to VUV light (160--650 nm spectral response).

The PMTs detect both the prompt primary scintillation (S1) light, as well as the delayed electroluminescent (S2) light induced by the ionization charge reaching the EL region. The time difference between the two light pulses defines the drift coordinate of the event, provided that the electron drift velocity is known. The detector can additionally be instrumented with a second readout plane. When present, this readout plane is placed a few mm behind the ground mesh, to provide better event position reconstruction in a plane perpendicular to the drift direction. Two types of anode readout planes have been tested in NEXT-DEMO: one based on 19 Hamamatsu R7378A PMTs, mirroring the cathode readout plane, and another one based on 256 Hamamatsu S10362-33-050C silicon photomultipliers (see \cite{nextdemopaper} for details).

The fast signals produced by the PMTs are first amplified, then shaped and filtered via a passive RC filter with a cut frequency of 800 kHz \cite{Gil:2012sr}, in order to better match the digitizer sampling rate and to eliminate high-frequency noise. The analog front-end circuit for NEXT-DEMO is connected via HDMI cables to 12-bit, 40-MHz, digitizer cards. These digitizers are read out by FPGA-based data-acquisition (DAQ) modules called front-end concentrator cards. Both the digitizer and the front-end concentrator cards have been designed in the framework of the scalable readout system for the RD51 R\&D program \cite{Martoiu:2013aca}. An additional DAQ module with a different plug-in card is used as trigger module. Besides forwarding a common clock and commands to all DAQ modules, this card receives trigger candidates from the DAQ modules, runs a highly-configurable trigger algorithm in the FPGA and distributes a global trigger signal. The trigger electronics accepts also external triggers for detector calibration purposes. Whenever a trigger is generated, a PC farm running the DAQ software, DATE, receives event data from the DAQ modules via Gigabit Ethernet (GbE) links. The DATE PCs assemble incoming fragments into sub-events, which are sent to one or more additional PCs (Global Data Concentrators, GDC). The GDCs build complete events and store them to disk for offline analysis. 

\subsection{Detector configuration} \label{subsec:ExperimentalSetupConfiguration}

\begin{figure}[t!b!]
\begin{center}
\includegraphics[scale=.30]{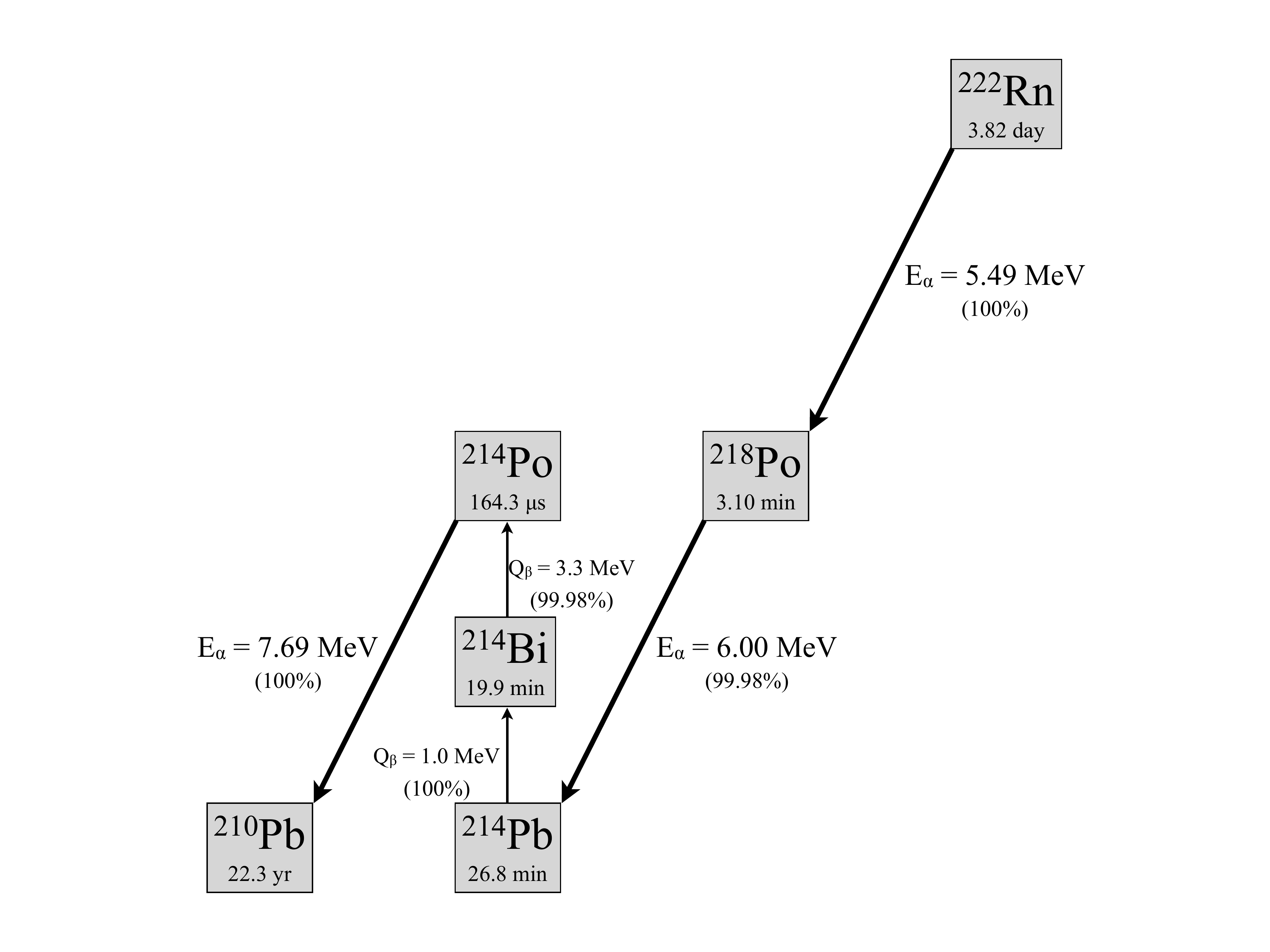}
\end{center}
\caption{The part of the $^{222}$Rn decay scheme showing the three alpha decays that are relevant for this analysis.}\label{fig:rn222decayscheme}
\end{figure}

All data discussed in this paper were taken during July 23-27, 2012. The detector was filled with pure xenon gas at 10 bar pressure. Gas purification was accomplished via a room-temperature SAES MC50 getter. For this NEXT-DEMO run, a flow-through radioactive source containing dry radium ($^{226}$Ra) powder was connected to the gas system. The source, manufactured by Pylon (model RNC), has an activity of approximately 74 Bq. The alpha decay of $^{226}$Ra produces $^{222}$Rn gas that diffuses through the entire detector. As a result, and as shown in Fig.~\ref{fig:rn222decayscheme}, three main sources of alpha particles are expected within NEXT-DEMO, from the decay of the $^{222}$Rn, $^{218}$Po and $^{214}$Po isotopes, respectively. The energy deposition of alpha particles is expected to be point-like in the detector. More precisely, the continuous slowing down approximation (CSDA) range of 5.49 MeV alpha particles from $^{222}$Rn in xenon at 10 bar pressure ($\rho = 5.56\cdot 10^{-2}$ g/cm$^3$ density) is only 2.4 mm. Based on previous findings (see, for example, \cite{Argyriades:2009vq}), we expect a different spatial distribution of alpha events depending on the parent nuclide. Radon is a neutral gas, therefore $^{222}$Rn decays are expected to occur throughout the detector. On the other hand, $^{218}$Po and $^{214}$Po daughters are mostly produced as positively charged ions, of which most of them are expected to accumulate on the negatively charged field grids and on the PTFE reflector panels before they are neutralized.

The NEXT-DEMO detector has been operated with PTFE reflectors both coated with terphenyl-butadiene (TPB) and uncoated, see \cite{nextdemopaper}. The goal of TPB is to shift the xenon VUV light ($\sim 172$ nm) into blue light ($\sim 430$ nm). This allows for a better light collection efficiency, thanks to the increased NEXT-DEMO reflectivity as well as the larger PMT spectral response.  
The data presented here were taken with TPB-coated PTFE reflectors. At the time of this run, the PMT-based anode readout plane had been dismantled, while the SiPM-based one was still under commissioning and not fully operational. For this reason, no anode readout plane information is used in this analysis. 

As for the field configuration, the gate voltage was kept at $-5$ kV throughout the alpha run, resulting in a reduced field between the EL meshes of $E_{\textrm{el}}/P\simeq 1.06$ kV/(cm$\cdot$bar). This field corresponds to a low EL gain of about 200 photons per ionization electron \cite{Oliveira:2011xk}. This near-threshold EL gain was chosen in order not to exceed the dynamic range of the digitizers, while maintaining the PMT at their nominal gain of about $5\cdot 10^6$ \cite{nextdemopaper}. Data were taken at three different drift field strengths, $E_{\textrm{drift}}=$0.3, 0.6, and 1 kV/cm, by supplying the cathode with a 14, 23, and 35 kV negative high voltage, respectively. 

Data were acquired by triggering on the ionization (S2) signal. The ionization signal charge was required to be greater than about 2,000 photoelectrons (PEs) per PMT, in order to reject low-energy events. The S2 signal was also required to have an amplitude not exceeding the digitizer dynamic range, and a time width of less than 20 $\mu$s. The latter requirement ensures keeping all alpha events, characterized by one sigma widths of the S2 pulses of 3 $\mu$s or less, while rejecting tracks with a large ($\gtrsim 2$ cm) extent along the drift direction, such as most cosmic ray muons. For each trigger, the individual PMT waveforms are stored over a 450--600 $\mu$s time interval. The DAQ time window was configured in order to have 100 $\mu$s worth of post-trigger data, and pre-trigger data amounting to 350--500 $\mu$s, depending on the drift field configuration. This time window ensured always recording the S1 signal corresponding to the triggered S2 signal, as it corresponds to more than one full drift length in NEXT-DEMO. The average trigger rate was about 5 Hz. Data were acquired for about 24 hours in each of the three drift field configurations.

\section{Data processing and selection} \label{sec:ProcessingSelection}

\subsection{Waveform treatment} \label{subsec:ProcessingSelectionWaveform}

A pre-processing of the raw data is performed in order to identify the S1-like and S2-like signals present in the individual channel waveforms, and to zero-suppress the data. First, pedestals are subtracted for all 19 PMT channels. The pedestal level is computed on a channel-by-channel and on an event-by-event basis, using the charge samples within the first 10,000 samples in the waveform that are not considered outliers. Second, pedestal-subtracted charge values are converted from ADC counts into photo-electrons. These (channel-dependent) gain calibration constants are obtained from single-PE LED spectra \cite{nextdemopaper} as well as from alpha events discussed below. Third, the 19 pedestal-subtracted, gain-corrected, PMT waveforms are summed together. This summed waveform is less affected by (uncorrelated) noise and by statistical fluctuations with respect to the individual ones of each PMT channel. For this reason, a peak-finding algorithm to isolate the S1 and S2 signals in the waveform is applied to this channel sum only, in a fourth data processing step.

\begin{figure}[t!b!]
\begin{center}
\includegraphics[scale=.60]{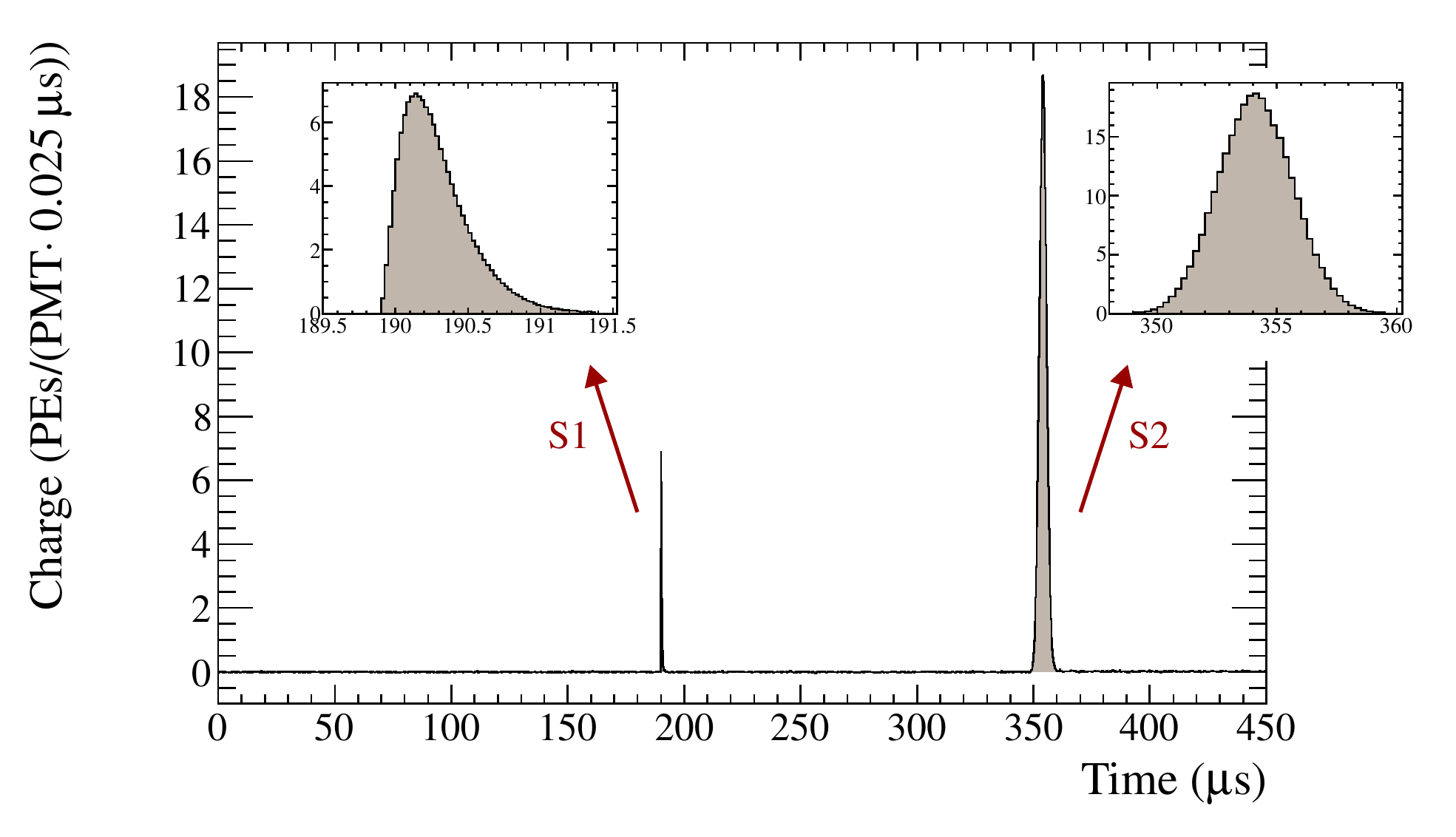}
\end{center}
\caption{Typical waveform of an alpha candidate event. This waveform is obtained by summing the pedestal-subtracted, gain-corrected, 19 individual PMT waveforms. The S1 and S2 signals, appearing also in the insets, are clearly visible. Alpha candidate events are characterized by intense S1 and S2 signals, and by fast (few $\mu$s long) S2 pulses.}\label{fig:waveform}
\end{figure}

A typical summed waveform for an alpha candidate event, as defined below in Sec.~\ref{subsec:ProcessingSelectionGasAlphas}, is given in Fig.~\ref{fig:waveform}. The (early) S1 and (late) S2 pulses can be clearly identified. The peak finding algorithm first builds pulses, or ``peaks'', as collections of consecutive samples above pedestal. Peaks can qualify as S1-like or S2-like depending on their time and charge characteristics. An S1-like peak is required to be less than $3~\mu$s wide, and to have an average charge per PMT of at least 0.5 PEs/PMT. An S2-like peak is defined as being at least $3~\mu$s wide, and with a charge of at least 500 PEs/PMT. Because of their wide pulse structure, for S2 peaks only, the charge samples are added together into $0.25~\mu$s time bins (that is, by applying a resampling factor of 10). This allows for some data reduction at no significant information loss. The bulk of the raw data reduction is obtained by dropping from the individual channel waveforms all samples that are not coincident in time with the S1-like and S2-like peaks.

\subsection{Selection of alpha-like ionization signals} \label{subsec:ProcessingSelectionPointlike}

The first step in the event selection, common to both alpha samples discussed in the following (Secs.~\ref{subsec:ProcessingSelectionCathodeAlphas} and \ref{subsec:ProcessingSelectionGasAlphas}), involves the analysis of the S2 signal only. We first require a single S2-like peak (as defined in Sec.~\ref{subsec:ProcessingSelectionWaveform}) in the entire channel sum waveform of the event. This requirement is applied to suppress complicated topologies, such as the ones involving accidental coincidences. We then require the S2 peak to be \emph{alpha-like} in order to select the event. An S2 peak is defined as alpha-like if it fulfills two conditions. First, the charge detected in time coincidence with the S2 peak must exceed 1,000 PEs for all of the 19 PMTs in the plane. This is to ensure that the S2 peak on the channel sum waveform is a genuine S2 signal, illuminating all PMTs at once, as opposed to uncorrelated noise on only some of the PMTs. Second, reinforcing what already specified at the trigger level, the S2 peak is required to have a narrow time width comprised between 2 and 18 $\mu$s, where the width is defined in this case as the full width at 5\% of the S2 pulse height. In other words, only energy depositions localized in 3D space (such as alpha decays) or extending only perpendicularly to the drift direction are expected to fulfill this requirement. A data reduction summary for the selection of alpha-like ionization signals is given in Tab.~\ref{tab:alphalikes2}.

\begin{table}[t!b!]
\caption{Data reduction summary for the selection of alpha-like ionization signals. The number of events after each selection requirement is given, for all three drift field configurations studied.}\label{tab:alphalikes2}
\begin{center}
\begin{tabular}{c|rrr}
\hline
Requirement      & \multicolumn{3}{c}{Drift Field (kV/cm)} \\
                 & 0.3     & 0.6     & 1.0  \\ \hline
None             & 361,796 & 488,017 & 397,391 \\
1 S2             & 347,103 & 478,203 & 368,974 \\
1 alpha-like S2  & 341,981 & 431,052 & 329,226 \\
\hline
\end{tabular}
\end{center}
\end{table}

\subsection{Selection of alpha decays from the cathode plane} \label{subsec:ProcessingSelectionCathodeAlphas}

A clean sample of alpha candidate events originating from the cathode plane (\emph{cathode alphas}, in the following) can be obtained by starting from the alpha-like ionization signal selection (Sec.~\ref{subsec:ProcessingSelectionPointlike}), and by imposing additional requirements on the scintillation signal. Such events are used to measure the electron drift velocity, see Sec.~\ref{subsec:ElectronTransportVelocity}. We first require at least one S1-like peak (as defined in sec.~\ref{subsec:ProcessingSelectionWaveform}) on the channel sum waveform. At this stage we allow for more than one S1 peak to avoid large selection efficiency losses, given that multiple spurious S1 peaks are often found by our peak finding algorithm. Among those peaks, we retain only the ones with a $<200$ ns rise-time (10\%--90\%) and an associated charge on each of the 19 PMTs of at least 200 PEs (\emph{cathode-like} peaks, in the following). The rise-time requirement eliminates S2-like pulse shapes, which are more symmetric and with longer rise-times compared to S1-like pulse shapes (see insets in Fig.~\ref{fig:waveform}). We define as cathode alpha candidates the events with one and only one S1 peak fulfilling the two conditions above. 

\begin{figure}[t!]
\begin{center}
\includegraphics[scale=.50]{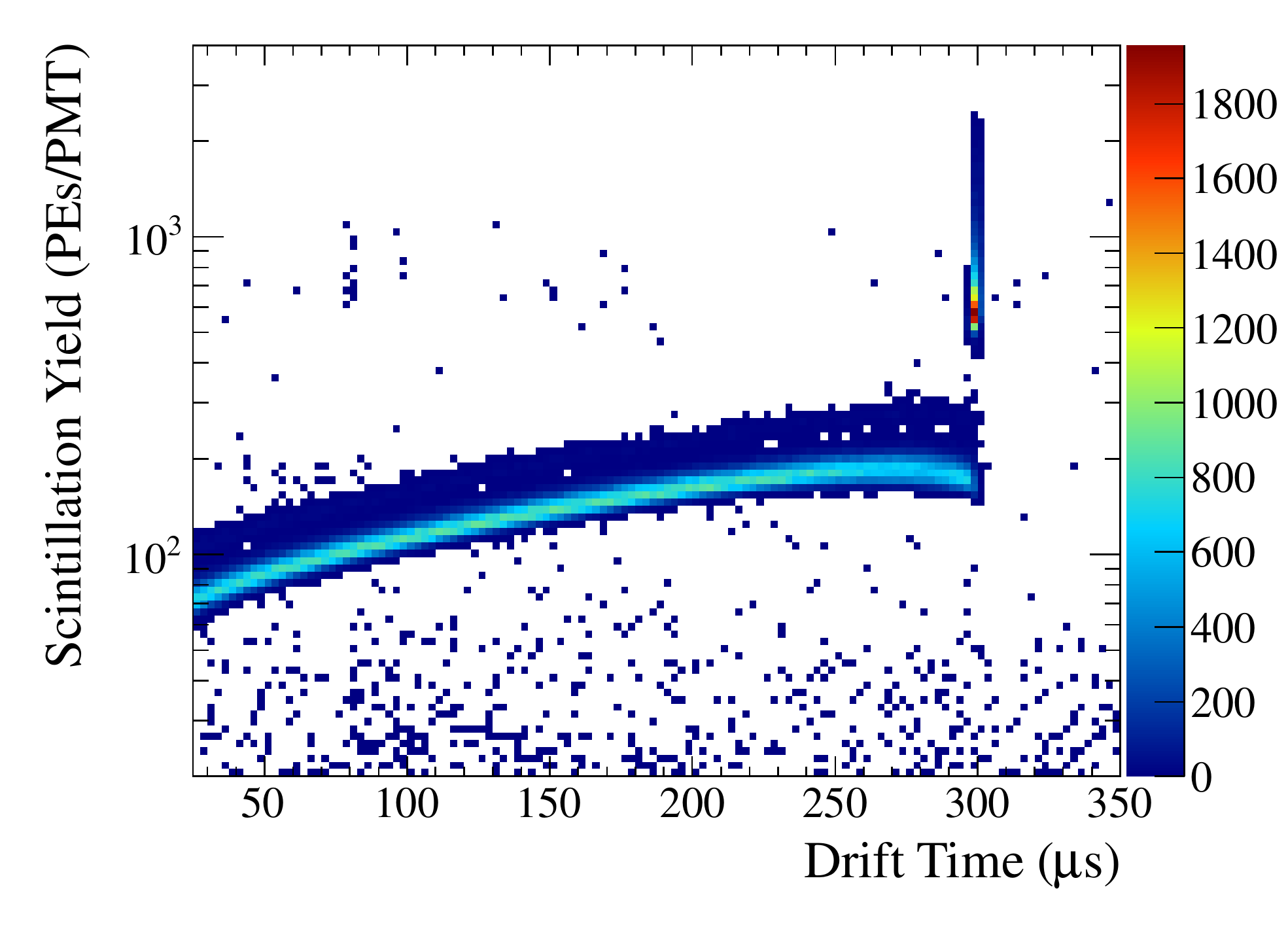}
\end{center}
\caption{Scintillation (S1) yield versus drift time for events in the $E_{\textrm{drift}}=0.6$ kV/cm run containing a single S1-like peak with $<200$ ns rise-time. Two populations, corresponding to alpha decays from the cathode plane and in the gas bulk, respectively, are clearly visible. The total number of detected scintillation photons is obtained by multiplying the yields shown in the figure by 19, the number of active PMTs.}\label{fig:cathodeselection}
\end{figure}

Cathode alpha events are characterized by abnormally large S1 signals, above the level expected for S1 signals from alpha decays in the gas, as shown in Fig.~\ref{fig:cathodeselection}. The figure shows the event S1 yield versus drift time for events prior to the $>$200 PEs charge per PMT requirement. The drift time is given by the difference between the (late) S2 and (early) S1 peak positions. The large scintillation signals at a fixed and long drift time (about 300 $\mu$s in the figure) correspond to alpha decays from the cathode plane. The band of 70--200 PEs/PMT scintillation signals for shorter drift times correspond to alpha decays in the gas bulk (see Sec.~\ref{subsec:ProcessingSelectionSpatialCorrections} for the dependence of the S1 yield with drift time). In addition, we have observed that the average S1 yield for the cathode alpha candidate events increase markedly with the high voltage supplied to the cathode, from $\sim 300$~PEs/PMT at $-14$ kV to $\sim 1,500$~PEs/PMT at $-35$ kV. The large S1 yields of cathode alpha candidate events are due to electroluminescence produced near the cathode wires. On the other hand, the S2 yield for such events is similar to that for alpha particles in the gas. In this case, the S2 charge distribution exhibits a bimodal distribution which is consistent with being due to $^{218}$Po (6.00 MeV) and $^{214}$Po (7.69 MeV) alpha decays. An event selection summary for cathode alpha candidate events is given in Tab.~\ref{tab:cathodealphas}

\begin{table}[t!b!]
\caption{Data reduction summary for the selection of alpha decays on the cathode plane. The number of events surviving each selection requirement is given, for all three drift field configurations studied.}\label{tab:cathodealphas}
\begin{center}
\begin{tabular}{c|rrr}
\hline
Requirement       & \multicolumn{3}{c}{Drift Field (kV/cm)} \\
                  & 0.3     & 0.6     & 1.0  \\ \hline
1 alpha-like S2   & 341,981 & 431,052 & 329,226 \\
$\ge$1 S1         & 287,797 & 357,797 & 262,916 \\ 
1 cathode-like S1 &  25,829 &  50,363 &  33,327 \\
\hline
\end{tabular}
\end{center}
\end{table}

\subsection{Selection of alpha decays in the gas bulk} \label{subsec:ProcessingSelectionGasAlphas}

A sample of alpha decays occurring in the xenon gas within the TPC (\emph{bulk alphas}, in the following) can also be isolated. Bulk alpha candidate events are used to measure longitudinal diffusion of ionization electrons (Sec.~\ref{subsec:ElectronTransportDiffusion}) as well as to perform electron-ion recombination studies (Sec.~\ref{sec:Recombination}). Bulk alpha events can be obtained by starting from the same sample of events with alpha-like ionization signals used above for the cathode alpha event selection. The following additional criteria are imposed. First, at least one S1-like peak is required. Second, we require exactly one pulse, among this pool of S1 candidate pulses per event, to be \emph{alpha-like}. An alpha-like S1 signal is defined as one having a signal rise-time of less than 200 ns, and 19 out of 19 PMTs in the readout plane measuring a charge of at least 20 PEs in time coincidence with the S1 peak. The PMT charge multiplicity requirement, less strict than for cathode-like alphas, is sufficient to ensure that genuine alpha-induced S1 signals are selected as opposed to noise (See Fig.~\ref{fig:cathodeselection}). 

\begin{figure}[t!]
\begin{center}
\includegraphics[scale=.50]{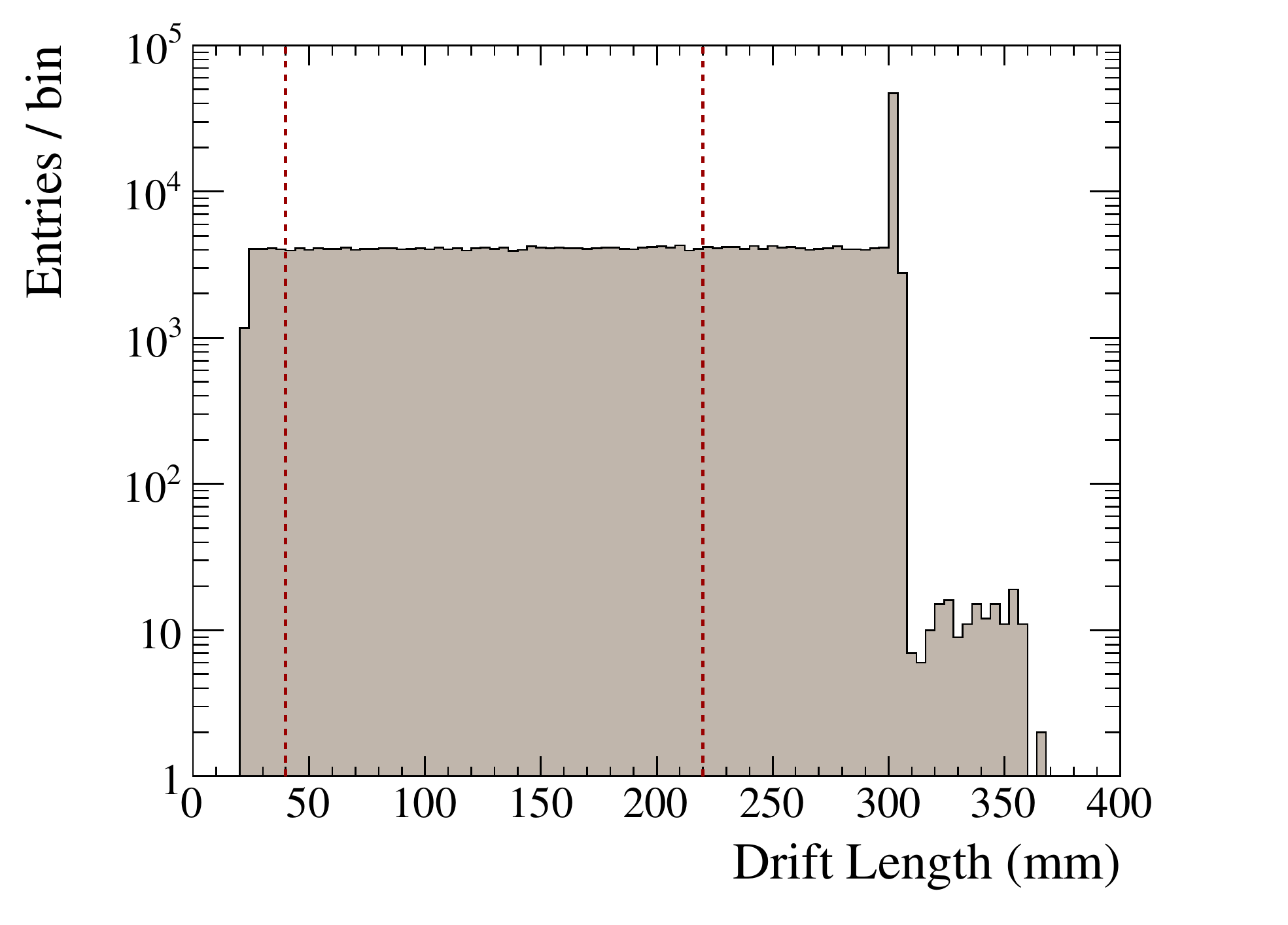}
\end{center}
\caption{Drift length distribution of events in the $E_{\textrm{drift}}=0.6$ kV/cm run, prior to the drift length cut. Events between the two vertical lines at 40 and 220 mm drift are retained. The large spike at $\sim$~300 mm drift corresponds to events originating from the cathode plane.}\label{fig:driftlengthcut}
\end{figure}

As a third condition, we require the event drift length $z$ to be comprised between 40 and 220 mm, as shown in Fig.~\ref{fig:driftlengthcut}. The drift length is the distance between the alpha decay location and the EL region, and is computed by multiplying the event drift time by the drift velocity measured for each drift field configuration, as discussed in Sec.~\ref{subsec:ElectronTransportVelocity}. As can be seen in Fig.~\ref{fig:driftlengthcut}, alpha decays occur throughout the full, 300 mm long, drift volume \footnote{The depletion of events at very small drift distances, $z\lesssim$~20 mm, is due to an artifact of the peak finding algorithm.}. The background level due to mis-reconstructed (S1,S2) pairs can be seen in the 300--350 mm drift length range, and is very small ($<1\%$). The large accumulation of events near 300 mm corresponds to alpha decays from the cathode plane, see Sec.~\ref{subsec:ProcessingSelectionCathodeAlphas}. Events at drift distances smaller than 40 mm are difficult to reconstruct because of the proximity of the S1 and S2 signals. We exclude events with drift distances between 220 mm and 300 mm because they involve a more complicated S1 yield spatial correction with respect to the linear one applied here, see Sec.~\ref{subsec:ProcessingSelectionSpatialCorrections}. 

\begin{figure}[t!]
\begin{center}
\includegraphics[scale=.37]{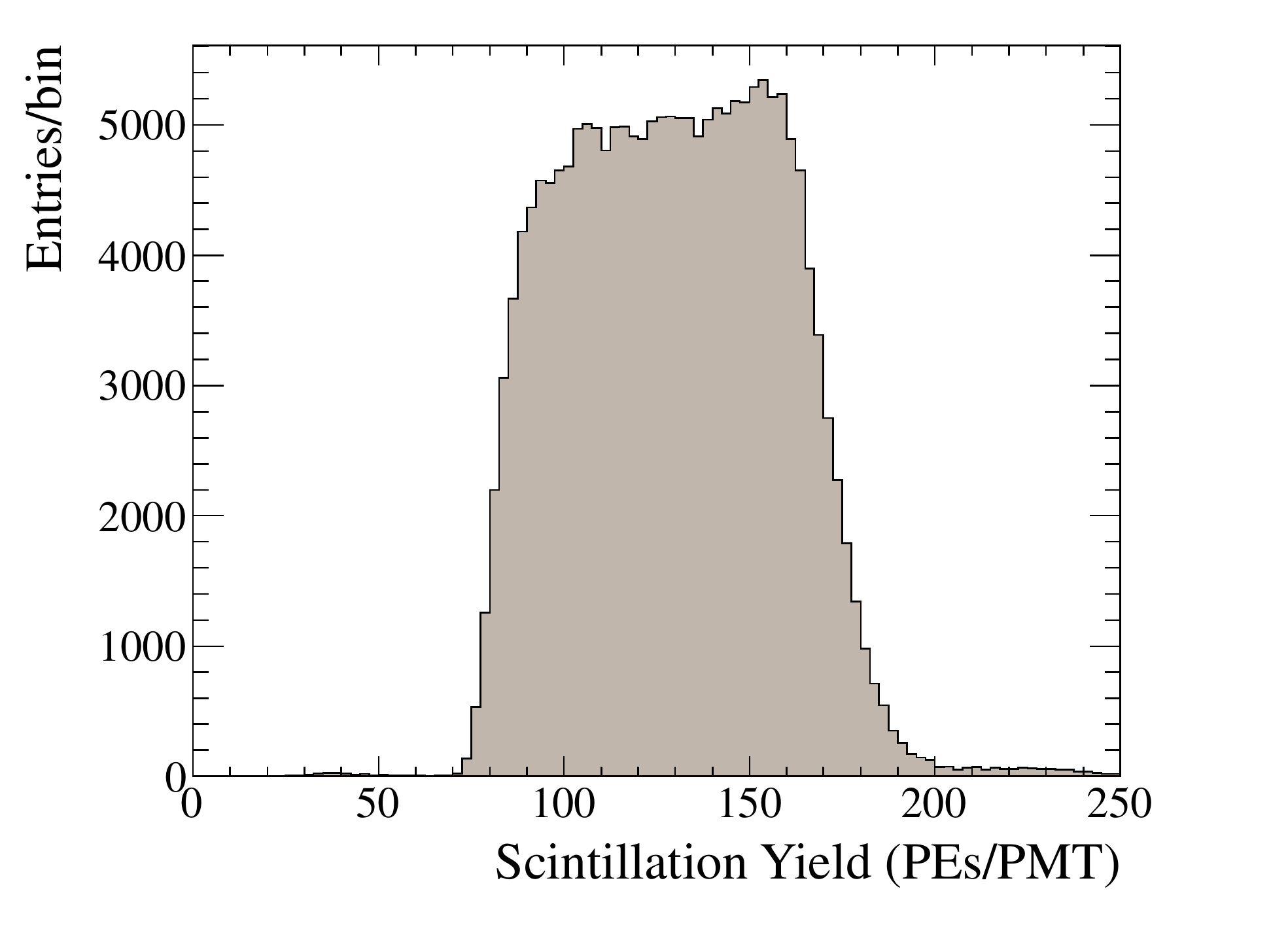} 
\includegraphics[scale=.37]{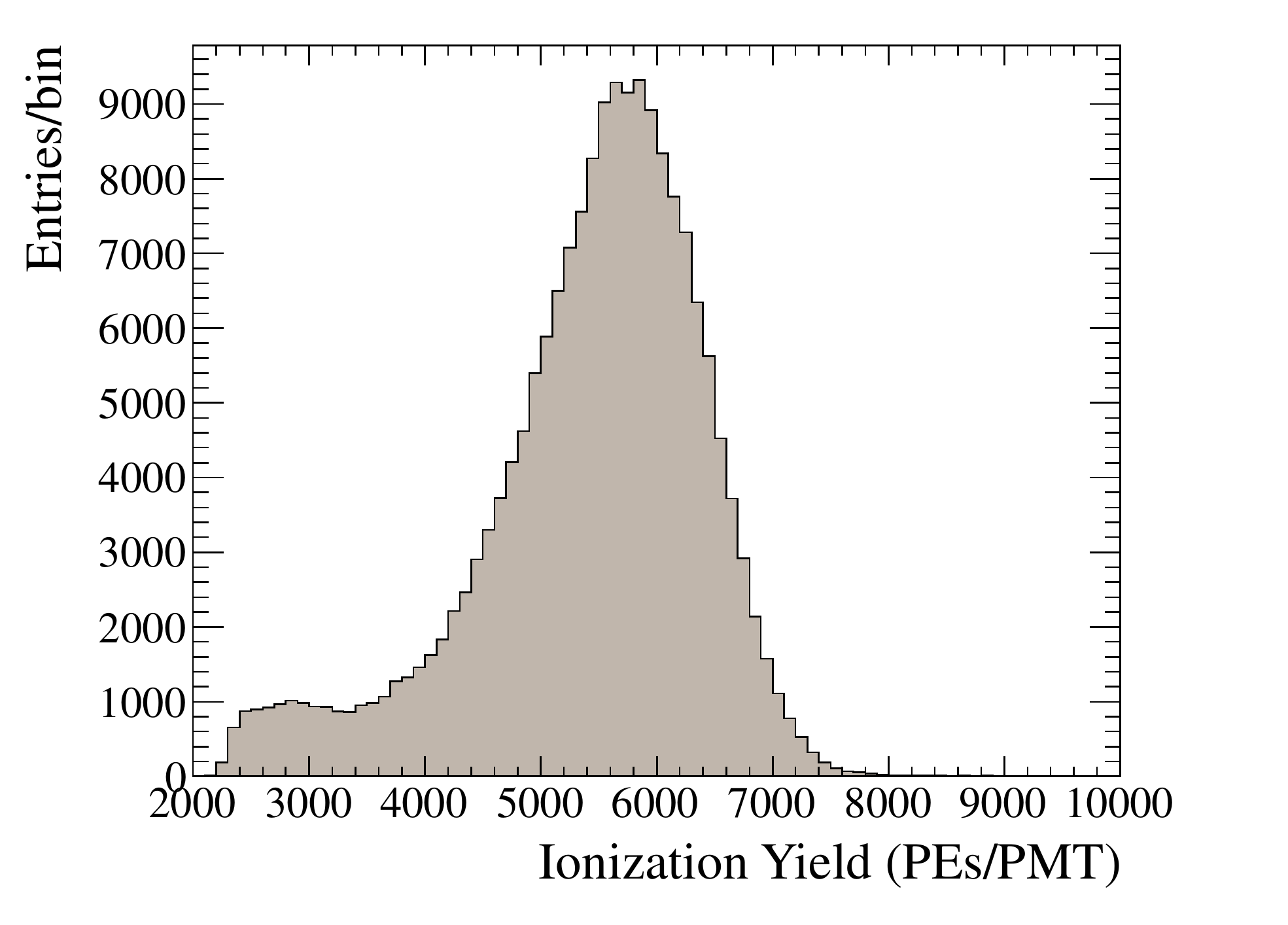}
\end{center}
\caption{Scintillation (S1, left panel) and ionization (S2, right panel) yield spectra for events in the $E_{\textrm{drift}}=0.6$ kV/cm run passing the drift length cut, with no corrections.}\label{fig:s1s2rawspectra}
\end{figure}

The S1 and S2 charge spectra for bulk alpha candidate events, after the drift length requirement and without any correction, are shown in Fig.~\ref{fig:s1s2rawspectra}. The width of the S1 distribution is due to the spatial dependence of the S1 signal, as discussed in Sec.~\ref{subsec:ProcessingSelectionSpatialCorrections}. The low-energy tail in the S2 distribution, which is present at all drift lengths and is well reproduced by our Monte Carlo simulations, is due to incomplete collection of the ionization charge at the electroluminescent region for alpha decays occurring at or near the PTFE reflector panels. Given that the transverse diffusion is $\sigma_z/\sqrt{z}\sim 1 \textrm{mm}/\sqrt{\textrm{cm}}$ for the gas pressure and drift field configurations under consideration, we expect to have sizable ionization charge losses for decays occurring at $\lesssim 10$ mm distances from the walls.

In order to reject alpha decays near (and at) the reflector walls, ideally the full 3-dimensional position of the decay location would need to be reconstructed. This can be best done with a second readout plane located near the EL region, not available for this analysis, as discussed in Sec.~\ref{sec:ExperimentalSetup}. As a consequence, only an S2 charge threshold cut was imposed in this analysis as a final event selection requirement. This cut suppresses alpha decays at large radius ($r\gtrsim 70$ mm) with respect to the TPC axis, by ensuring a (close to) complete collection of the ionization charge at the anode. In practice, a small amount of charge is always lost during drift because of electron attachment on electronegative impurities (see Sec.~\ref{subsec:ProcessingSelectionSpatialCorrections}), regardless of the decay $(x,y)$ position. For this reason, the S2 analysis threshold adopted here depends on the decay $z$ position, and corresponds to a minimum equivalent charge of 5,200 PEs/PMT at zero drift length. A summary of the data reduction for bulk alpha candidate events is given in Tab.~\ref{tab:gasalphas}.

\begin{table}[t!b!]
\caption{Data reduction summary for the selection of alpha decays in the gas bulk. The number of events surviving each selection requirement is given, for all three drift field configurations studied.}\label{tab:gasalphas}
\begin{center}
\begin{tabular}{c|rrr}
\hline
Requirement            & \multicolumn{3}{c}{Drift Field (kV/cm)} \\
                       & 0.3     & 0.6     & 1.0  \\ \hline
1 alpha-like S2        & 341,981 & 431,052 & 329,226 \\
$\ge$1 S1              & 287,797 & 357,797 & 262,916 \\ 
1 alpha-like S1        & 263,809 & 334,041 & 241,444 \\
Drift length cut       & 156,950 & 184,043 & 134,926 \\
S2 Charge Threshold    & 133,717 & 156,181 & 109,775 \\ 
\hline
\end{tabular}
\end{center}
\end{table}

\subsection{Spatial corrections to ionization and scintillation signals} \label{subsec:ProcessingSelectionSpatialCorrections}

The ionization and scintillation signals are affected by well-understood instrumental effects, causing the observed yields to depend on the alpha decay spatial position within the gas. Once events near (or at) the reflector walls are removed, the next most important (and easiest to correct) effect is given by the dependence of both the ionization and scintillation signals with drift length $z$. This correction is discussed below, using the sample of bulk alpha candidate events (see Sec.~\ref{subsec:ProcessingSelectionGasAlphas}). There is also a $(x,y)$ dependence of the yields, as discussed in \cite{nextdemopaper}. Given the lack of ($x,y)$ reconstruction capability in this analysis, no $(x,y)$, or radial, correction can be applied in this case.

\begin{figure}[t!b!]
\begin{center}
\includegraphics[scale=.50]{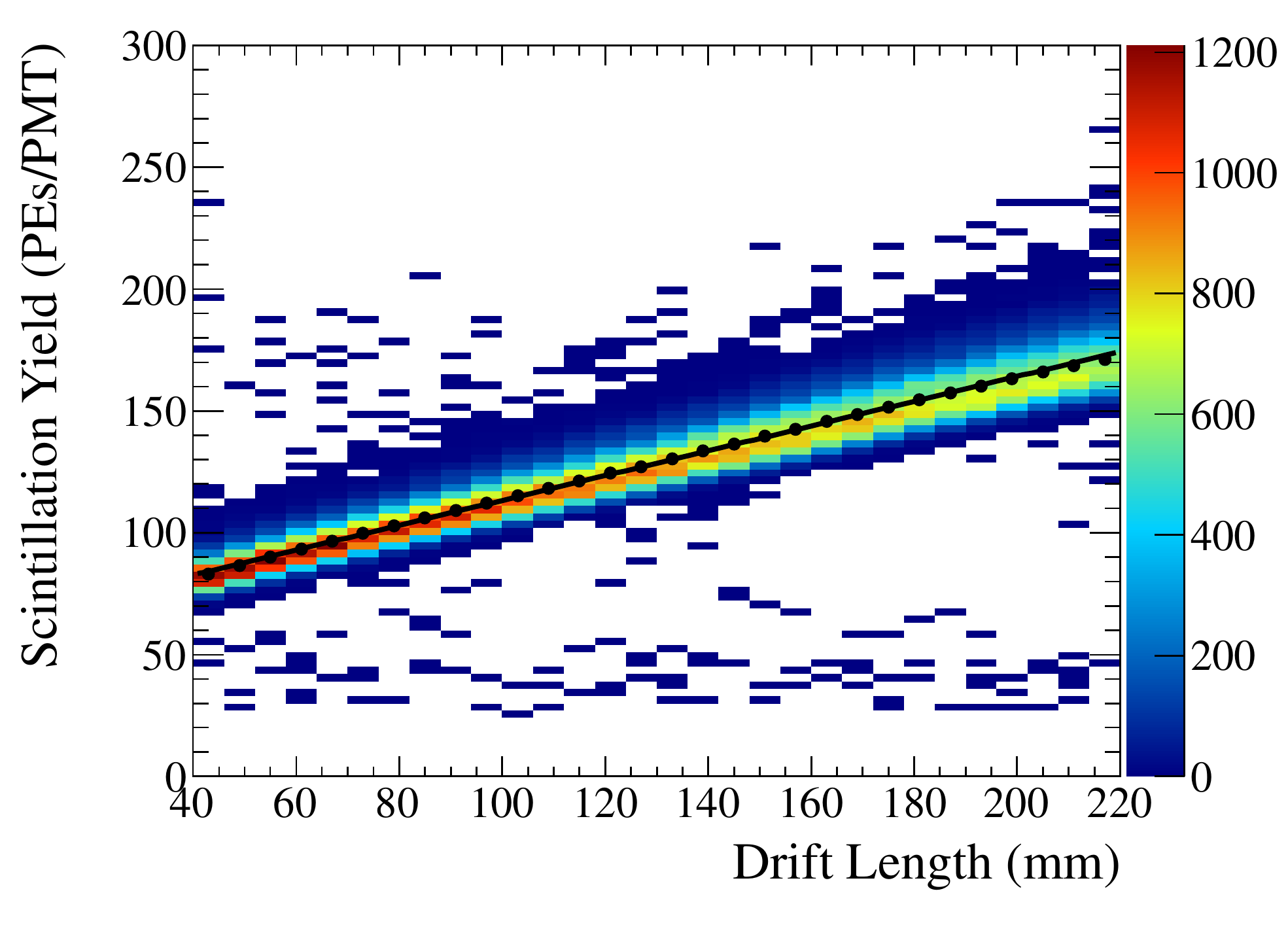}
\end{center}
\caption{Dependence of the S1 signal yield as a function of drift length. The 2D histogram of gas alpha candidate events for the $E_{\textrm{drift}}=0.6$ kV/cm run (colored boxes), the corresponding 1D histogram of S1 mean yields for each drift length bin (black points), and a linear fit to the 1D histogram (curve) are shown.}\label{fig:s1spatialcorr}
\end{figure}

The dependence of the scintillation signal with drift length is shown in Fig.~\ref{fig:s1spatialcorr}. As the alpha decay occurs closer to the energy plane located behind the cathode (large drift lengths), more scintillation light is detected on average. The main reason for this dependence has been understood via simulations as due to the non-perfect optical transparency of the gate and anode meshes, with an open area of 76\% and 88\%, respectively. The further away from the anode the alpha decay occurs, the smaller the fraction of scintillation light that cross the meshes and that can be absorbed. Given the relatively high ($\gtrsim 70\%$) reflectivity of the TPB-coated reflector panels under VUV and blue light, the TPB coating plays a non-negligible but less important role in understanding this spatial dependence. In particular, in the fiducial region chosen for bulk alphas event selection, between 40 and 220 mm drift length, the average S1 yield increases by about a factor of 2, increasing approximately linearly with drift length. This behavior is well described by simulations assuming the accurately known mesh transparencies, a VUV reflectivity of about 70\% (as determined by data with uncoated reflector panels), and a $\gtrsim 90\%$ reflectivity to blue light. The S1 yields shown in the following for recombination studies (Secs.~\ref{subsec:RecombinationFluctuations} and \ref{subsec:RecombinationFieldDependence}) are corrected according to this linear dependence, to provide an S1 yield at zero drift length. While the data in Fig.~\ref{fig:s1spatialcorr} refer to 0.6 kV/cm drift field data only, we have verified that this spatial dependence is independent of drift field strength, as expected from the above considerations. As mentioned above and as shown in Fig.~\ref{fig:cathodeselection}, the S1 dependence on drift length becomes milder for drift lengths larger than 220 mm, and a broad maximum is observed in the 250--300 mm range, for the case of TPB-coated reflector panels. This tendency has again been understood using simulations. Alpha decays occurring near the cathode exhibit less VUV-to-blue wavelength shifting of the photons reaching the PMTs, because of the higher fraction of direct (not reflected) light, and because the reflector panels in the buffer region have no TPB coating. Considering that the PMT quantum efficiency is lower for VUV ($\simeq 15\%$) than for blue ($\simeq 25\%$) light, this effect almost perfectly compensates for the reduced absorption of scintillation light by the meshes as one approaches the cathode.  

\begin{figure}[t!b!]
\begin{center}
\includegraphics[scale=.50]{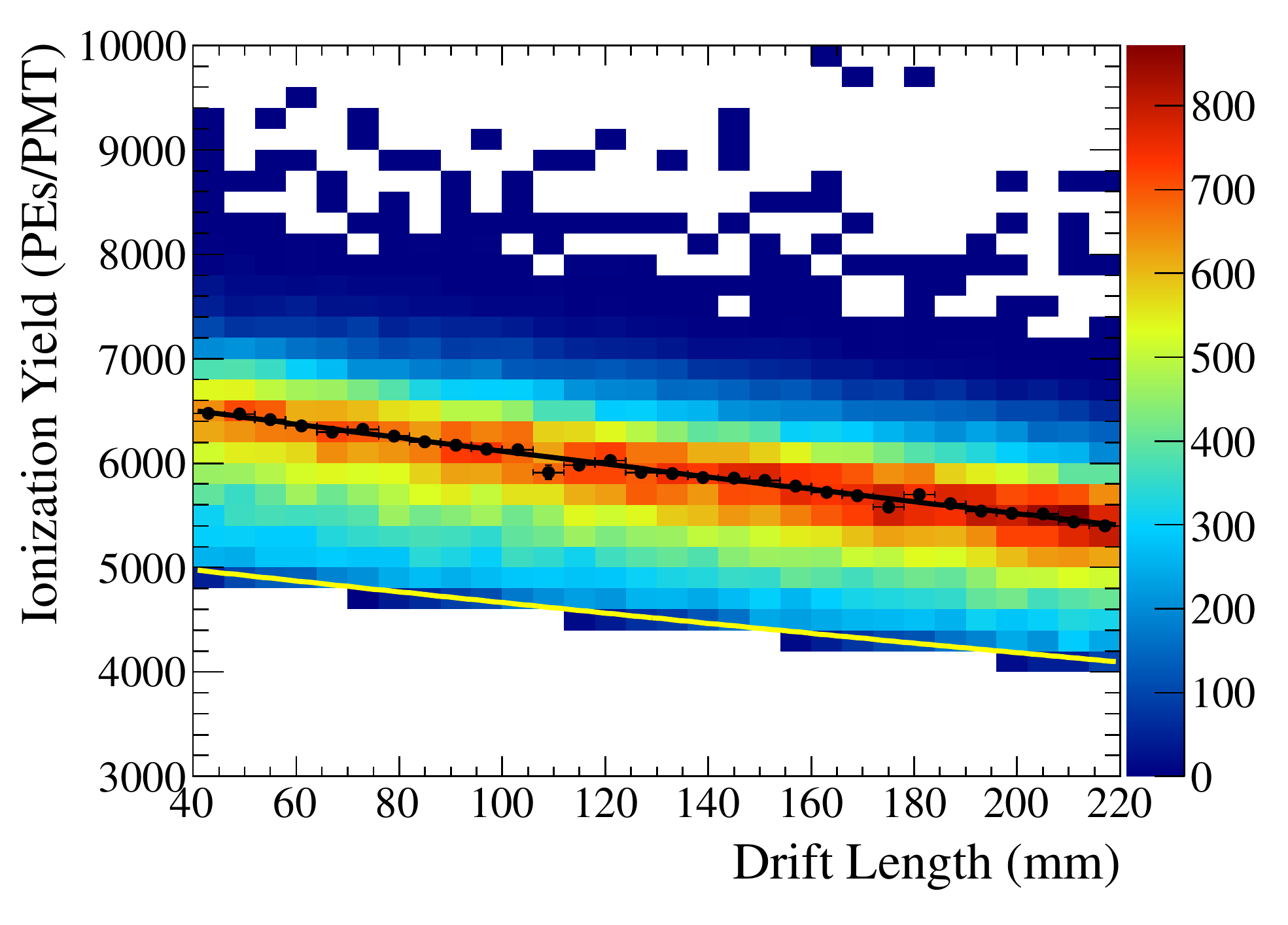}
\end{center}
\caption{Dependence of the S2 signal yield as a function of drift length. The 2D histogram of bulk alpha candidate events for the $E_{\textrm{drift}}=0.6$ kV/cm run (colored boxes), the corresponding 1D histogram of S2 peak positions for each drift length bin (black points), and an exponential fit to the 1D histogram (curve) are shown. The solid yellow line indicates the analysis S2 threshold.}\label{fig:s2spatialcorr}
\end{figure}

As shown in Fig.~\ref{fig:s2spatialcorr}, also the average ionization yield varies with drift length. In this case, the dependence is opposite, with more detected charge closer to the EL region (short drift lengths). Impurities in the xenon gas, causing some amount of attachment of ionization electrons, are responsible for this effect. The 0.6 kV/cm data shown in Fig.~\ref{fig:s2spatialcorr} are well described by an exponential dependence on $z$, corresponding to a $\sim$1 ms electron lifetime, and to a $\sim$20\% reduction in the average S2 yield between 40 and 220 mm drift length. Unlike for the S1 yield, this correction is dependent on the drift field configuration. This attachment effect is the largest in the lowest drift field configuration (0.3 kV/cm), where the average S2 variation across the 40--220 mm drift length range becomes $\sim$35\%. The S2 yields shown in Secs.~\ref{subsec:RecombinationFluctuations} and \ref{subsec:RecombinationFieldDependence} are corrected according to this exponential dependence, to provide an S2 yield at zero drift length. 

In order to minimize the dependence between the event selection for bulk alpha candidate events and the extraction of the attachment correction, the latter is computed with no S2 charge threshold cut. The S2 charge spectra for each drift length slice are fitted to an asymmetric gaussian distribution, to extract the best-fit value and error of the S2 peak position. The fitted peak positions versus drift length are also shown in Fig.~\ref{fig:s2spatialcorr}. We have verified that this procedure yields a consistent attachment correction even if computed from an event sample requiring the S2 analysis threshold.

\section{Measurement of ionization electron transport properties} \label{sec:ElectronTransport}


\subsection{Electron drift velocity} \label{subsec:ElectronTransportVelocity}

\begin{figure}[t!b!]
\begin{center}
\includegraphics[scale=.50]{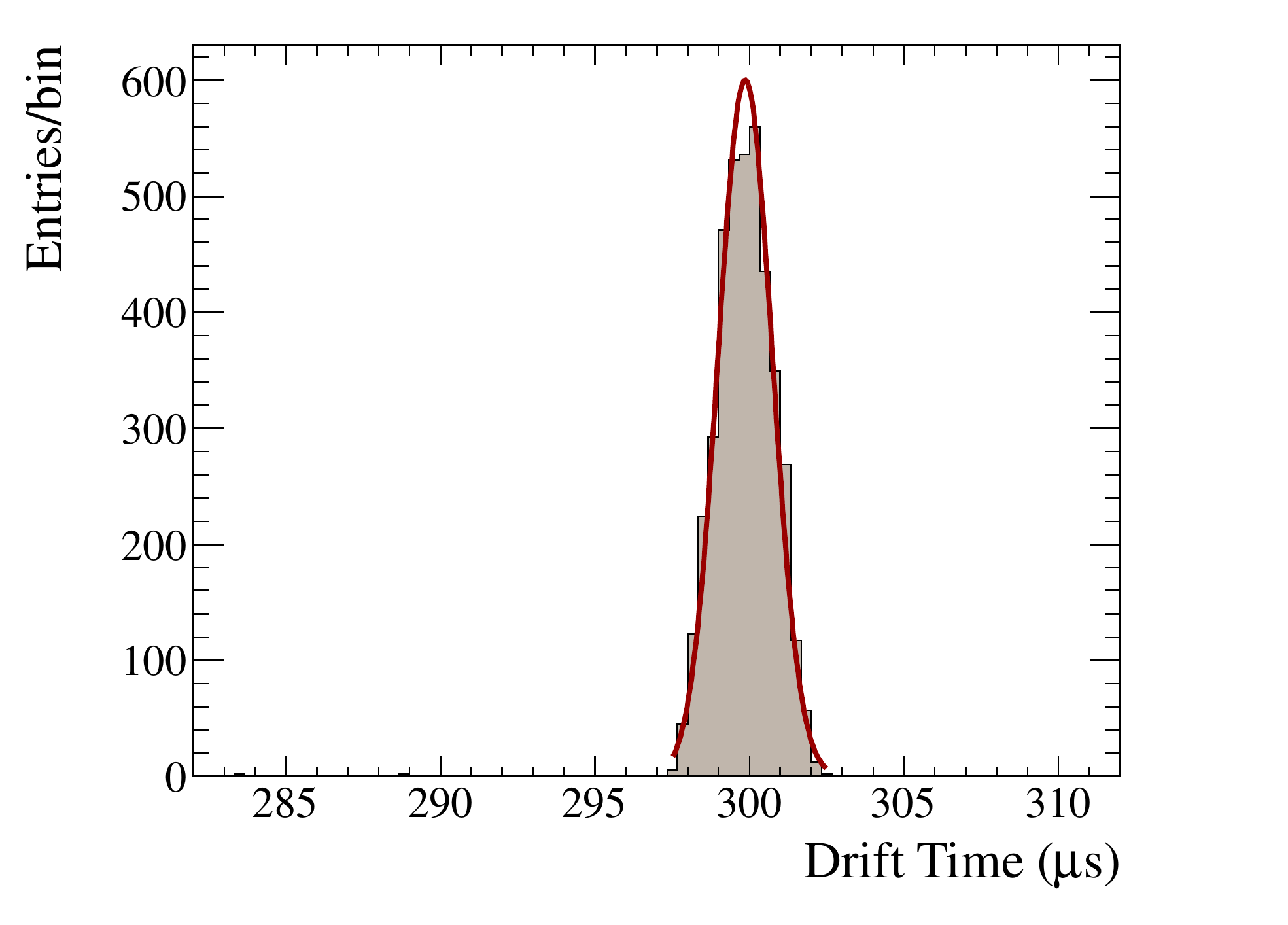}
\end{center}
\caption{Drift time for cathode alpha candidate events. Only data from approximately the first two hours of the $E_{\textrm{drift}}=0.6$ kV/cm run are shown. The red curve is a gaussian fit to the data.}\label{fig:cathodealphasdrift}
\end{figure}

The sample of cathode alpha candidate events (Sec.~\ref{subsec:ProcessingSelectionCathodeAlphas}) allows for an accurate determination of the electron drift velocity. A drift time distribution for such events is shown in Fig.~\ref{fig:cathodealphasdrift}. Specifically, the events shown in the figure refer to the first two hours of data taken in the $E_{\textrm{drift}}=0.6$ kV/cm configuration. The data are well described by a gaussian distribution (also shown in Fig.~\ref{fig:cathodealphasdrift}) with a mean drift time of $(299.848 \pm 0.023)~\mu$s and a one sigma spread of $0.9~\mu$s. Factors contributing to this (small) spread in drift times include noise, the relatively wide (few $\mu$s) ionization signal time extent, the orientation of the alpha decays with respect to the drift direction, sub-mm non-uniformities in the NEXT-DEMO full drift distance as a function of radial position and drift velocity variations with time (see below). 

We obtain a measurement of the drift velocity from the ratio between a fixed drift distance for all cathode alpha candidate events, and the mean drift time shown in Fig.~\ref{fig:cathodealphasdrift}. The relevant nominal distances are the 300 mm distance between the cathode and the gate grids, the 4.7 mm distance between the gate and the ground grids (the width of the EL region), and the 2.3 mm average range projected along the particle's initial direction for cathode alpha candidate events. Recalling that the drift time is given by the ionization signal peak time minus the scintillation signal peak time, we take $300+(4.7-2.3)/2=301.2$ mm as our best estimate for this fixed drift distance. Deviations as large as 2\% from this nominal value are considered below as a systematic uncertainty affecting absolute drift velocity measurements.

\begin{figure}[t!b!]
\begin{center}
\includegraphics[scale=.60]{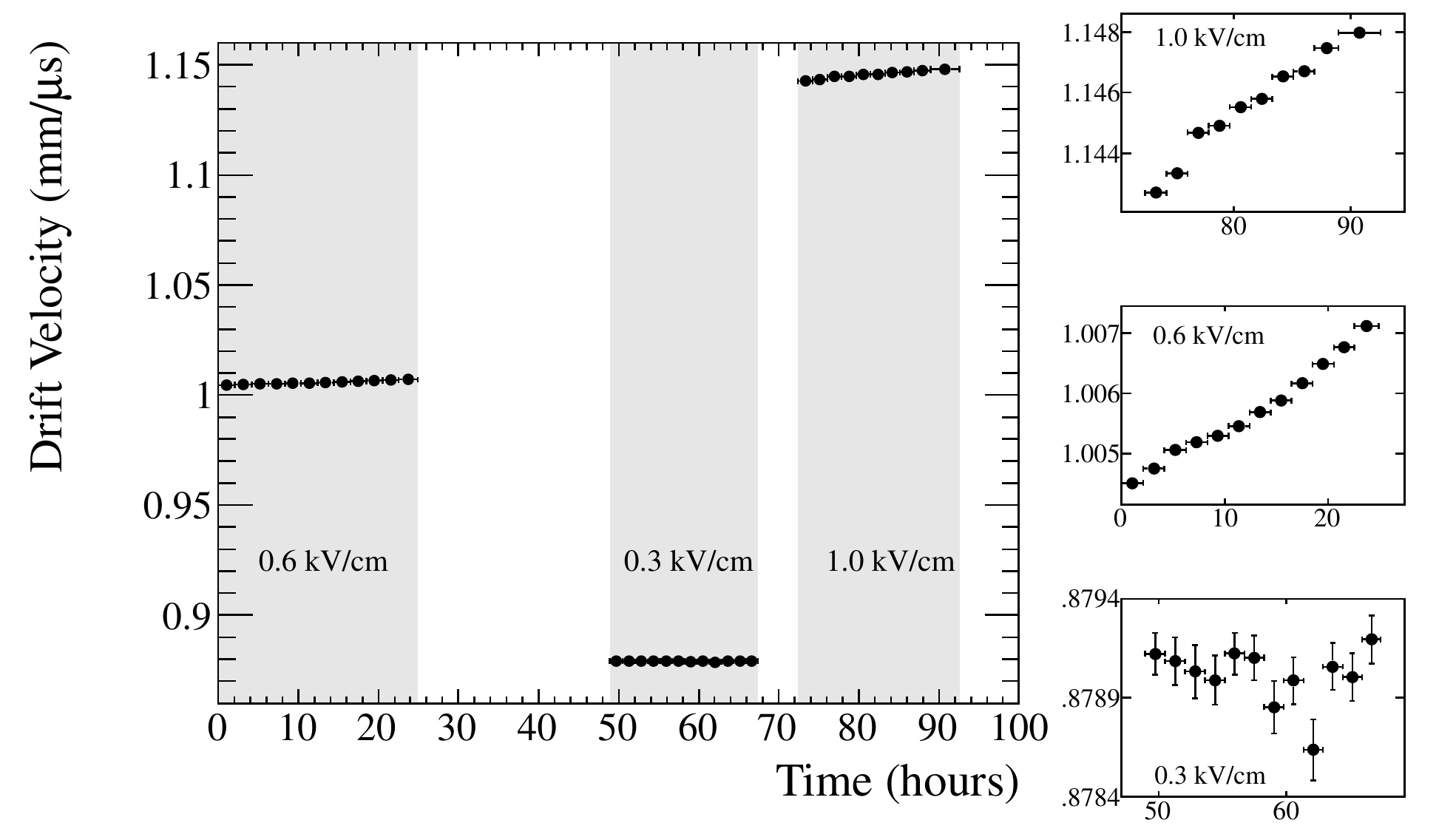}
\end{center}
\caption{Drift velocity as a function of time, for the three drift field configurations studied. For each time interval, the data points and errors shown are obtained from a gaussian fit to the drift time data as the one in Fig.~\protect\ref{fig:cathodealphasdrift}, plus the known full drift length of the NEXT-DEMO detector. The three small panels on the right show a zoomed version of the same drift velocity data.} \label{fig:driftvelocitystability}
\end{figure}

The variation in the drift velocity measurements as a function of time, and for approximately two hour time intervals, is shown in Fig.~\ref{fig:driftvelocitystability}. The main aspect to highlight is the excellent stability over time, with $<$0.5\% drift velocity variations during more than one day of data. Even so, the accuracy of the measurements is sufficient to appreciate a systematic trend in the data in the form of a very slow drift velocity increase over time, at least for the 0.6 and 1.0 kV/cm runs. This small increase in drift velocity over time can be explained by a gas leak that was observed (by manual inspection of the pressure gauges) during the run. The observed pressure leakage average rate was about 0.5\%/day \cite{nextdemopaper}. Gas was routinely (every few days) added to the chamber to compensate for the gas leak but, unfortunately, no detailed track record of the pressure conditions during the run exists. From the above considerations, in the following we will assume a 2\% systematic uncertainty on the pressure conditions in NEXT-DEMO during this run.

\begin{figure}[t!b!]
\begin{center}
\includegraphics[scale=.50]{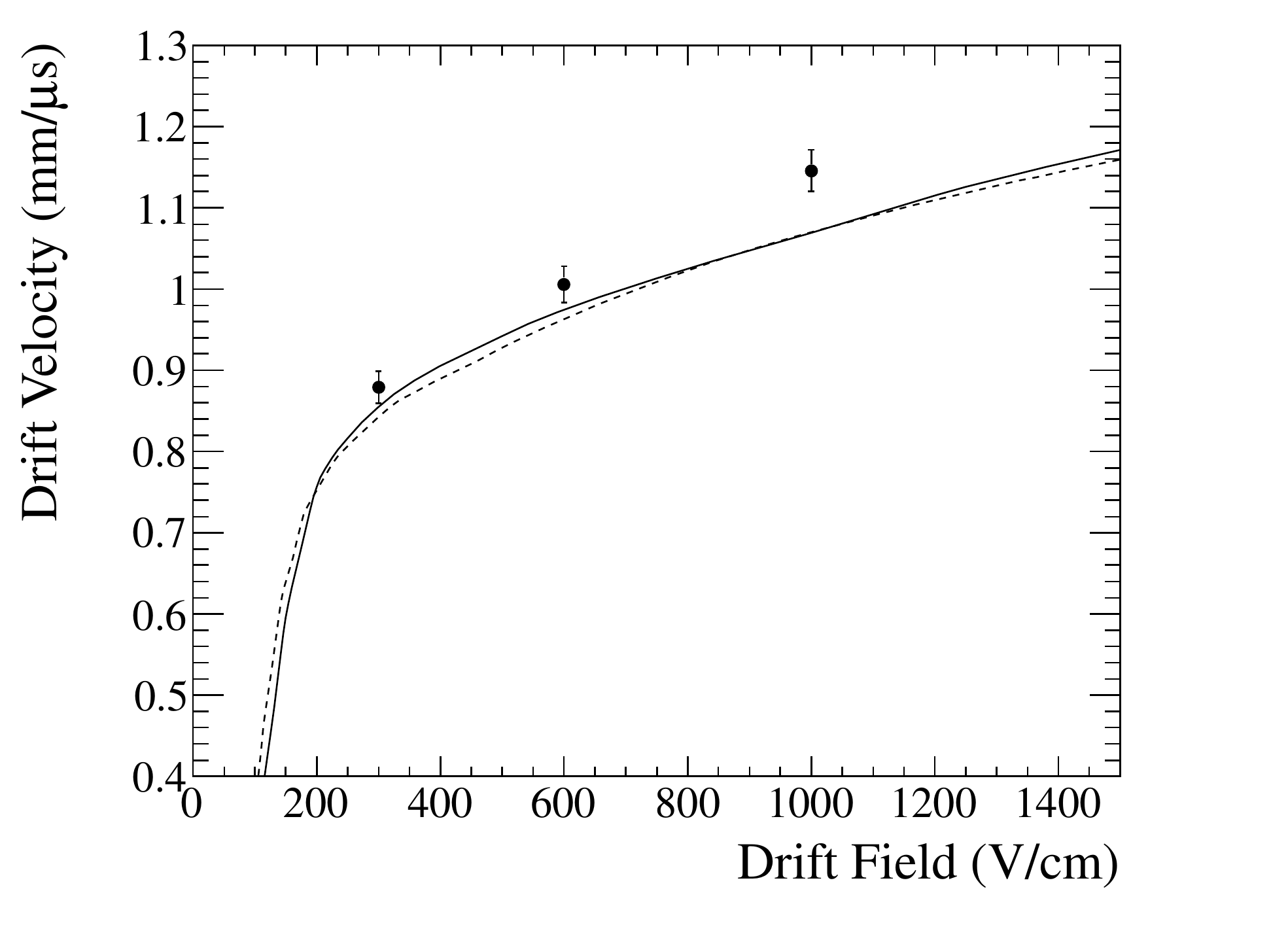}
\end{center}
\caption{Drift velocity as a function of drift field, for xenon gas at 10 bar pressure. The black points are the measured values, the solid (dashed) curve is the prediction for pure xenon from the Magboltz \protect\cite{magboltz} (Escada \textit{et al.} \protect\cite{Escada:2011cs}) simulation. The Magboltz results are based on version 9.0.1 of the code, and include thermal motion of atoms. The error bars on the NEXT-DEMO data account for the main uncertainties (drift length and pressure, see text for details).}\label{fig:driftvelocityvsfield}
\end{figure}

We have also compared our measured drift velocities at 0.3, 0.6, and 1 kV/cm drift fields with expectations from pure xenon gas at 10 bar pressure obtained with two independent simulations of electron transport in gases \cite{magboltz, Escada:2011cs}. The comparison is shown in Fig.~\ref{fig:driftvelocityvsfield}. The agreement is fair for all drift field configurations and for both simulations, with observed drift velocities only a few \% larger than the expected ones for pure xenon. The observed drift velocity central values were taken by simply averaging over the entire run at a given drift field. The error bars shown in Fig.~\ref{fig:driftvelocityvsfield} assume two main contributions, summed in quadrature: a 2\% uncertainty in the NEXT-DEMO full drift length (directly translating into a 2\% drift velocity uncertainty), and the 2\% pressure reading uncertainty mentioned above (translating into a $\sim$1\% drift velocity uncertainty through the associated change in $E_{\textrm{drift}}/P$ \cite{lxcat}).

\subsection{Electron longitudinal diffusion} \label{subsec:ElectronTransportDiffusion}

The diffusion of ionization electrons as a function of drift distance can be studied with bulk alpha candidate events (Sec.~\ref{subsec:ProcessingSelectionGasAlphas}), by exploiting the point-like nature of alpha particle energy depositions within the chamber. As mentioned in Sec.~\ref{subsec:ExperimentalSetupConfiguration}, the range of alpha particles is expected to be of the order of 2 mm, roughly of the same order of (and often smaller than) diffusion effects.

In this work, we infer the longitudinal diffusion of ionization electrons in xenon gas by studying how the ionization signal pulse width stretches with increasing drift length. In our simple method (see, for example, \cite{Cennini:1994ha,Sorensen:2011qs} for earlier works adopting essentially the same method), we first fit the S2 pulse shape to a gaussian distribution, in order to obtain the one sigma width $\sigma_t$ for each event. We then fit the entire data set of $\sigma_t$ values to the following function:

\begin{equation}
\sigma_{t}=\sqrt{\sigma_0^2+\sigma_L^2}
\label{eq:diffusionfit}
\end{equation}

\noindent where $\sigma_0$ is a constant term, and the longitudinal diffusion term $\sigma_L$, which varies with drift length, is defined as:

\begin{equation}
\sigma_L^2=(2D_L/v_d^3)\cdot z
\label{eq:sigmal}
\end{equation}

\noindent where $z$ is the drift length, $v_d$ is the electron drift velocity, and $D_L$ is the \textit{longitudinal diffusion coefficient} (in [cm$^2$/s] units). In our model, there are two free fit parameters, $\sigma_o$ and $D_L$, while $v_d$ is taken from our measured values using cathode alpha candidate events (Sec.~\ref{subsec:ElectronTransportVelocity}).

\begin{figure}[t!b!]
\begin{center}
\includegraphics[scale=.50]{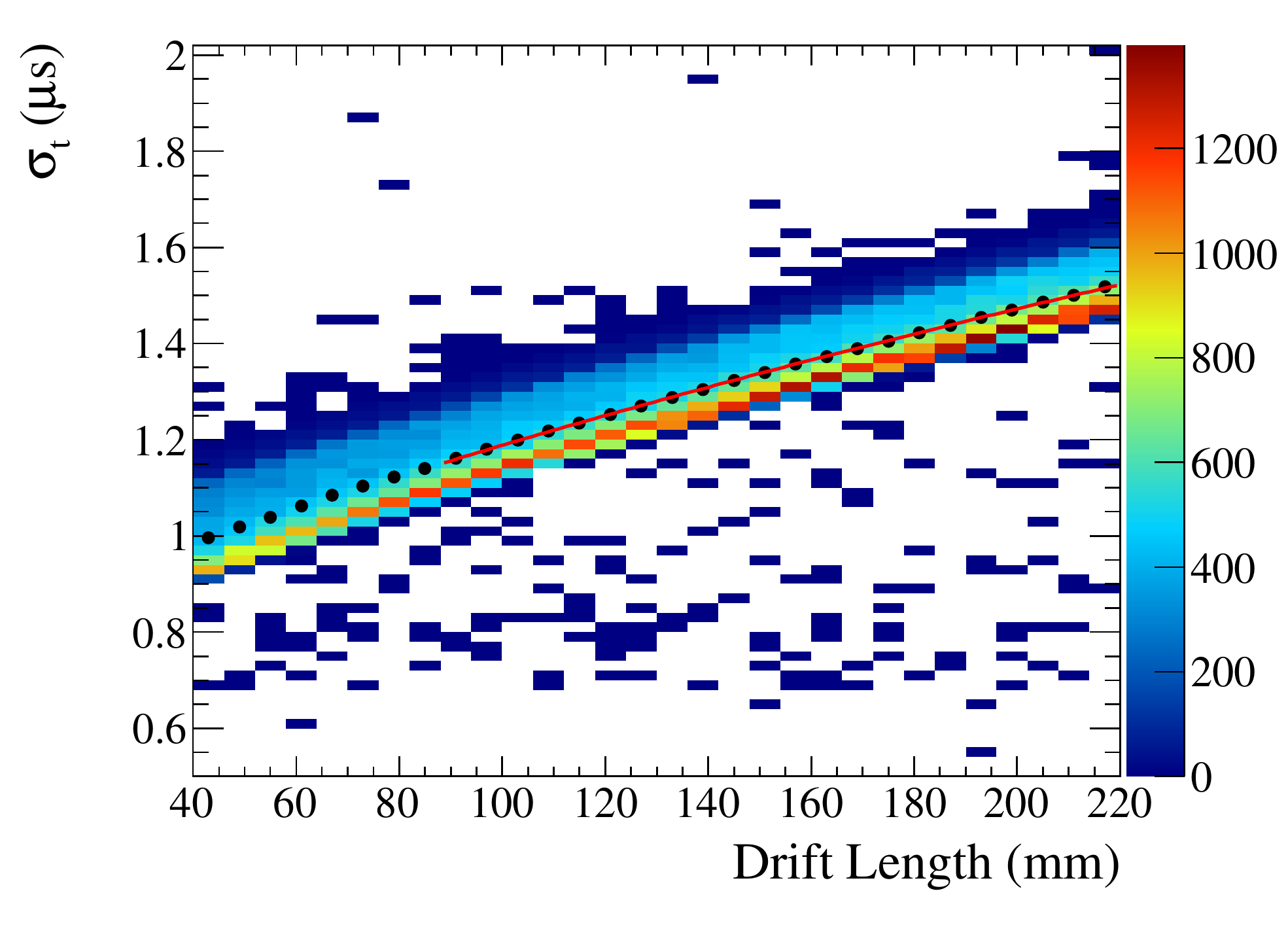}
\end{center}
\caption{Dependence of the S2 pulse time width ($\sigma_{t}$) as a function of drift length, illustrating the effect of longitudinal diffusion. The 2D histogram of bulk alpha candidate events for the $E_{\textrm{drift}}=0.6$ kV/cm run (colored boxes), the corresponding 1D histogram of mean $\sigma_{t}$ values for each drift length bin (black points), and a fit to the 1D histogram according to Eq.~\protect\ref{eq:diffusionfit} (curve) are shown.}\label{fig:diffusionfit}
\end{figure}

Figure \ref{fig:diffusionfit} shows the dependence of $\sigma_{t}$ in Eq.~\ref{eq:diffusionfit} with drift length $z$, for the 0.6 kV/cm drift field configuration. The 2D histogram of $\sigma_{t}$ versus drift length for all bulk alpha candidate events shows a clearly defined band, indicating a pure sample of correctly reconstructed alpha candidate events. In other words, few mis-reconstructed events due to accidental S1$+$S2 combinations, and few events with extended energy depositions along the drift direction, survive our event selection described in Sec.~\ref{subsec:ProcessingSelectionGasAlphas}. 

The 1D histogram of mean $\sigma_{t}$ values for each drift length bin, also shown in Fig.~\ref{fig:diffusionfit}, is used as input to our simple model given by Eq.~\ref{eq:diffusionfit}. The best-fit line resulting from the fit is also shown in Fig.~\ref{fig:diffusionfit}. Our simple model provides a fair but far from perfect description of the data at small drift lengths. For small drifts, diffusion effects do not fully dominate the S2 pulse shape, and the latter deviates from a gaussian distribution to follow a more square-like (that is, with negative excess kurtosis) pulse shape. The gaussian approximation to the S2 pulse shape is found to be sufficiently good for $\sigma_{t}>1.15\ \mu$s. Therefore, only the range of drift lengths satisfying this condition are used in the longitudinal diffusion fit.

\begin{figure}[t!b!]
\begin{center}
\includegraphics[scale=.50]{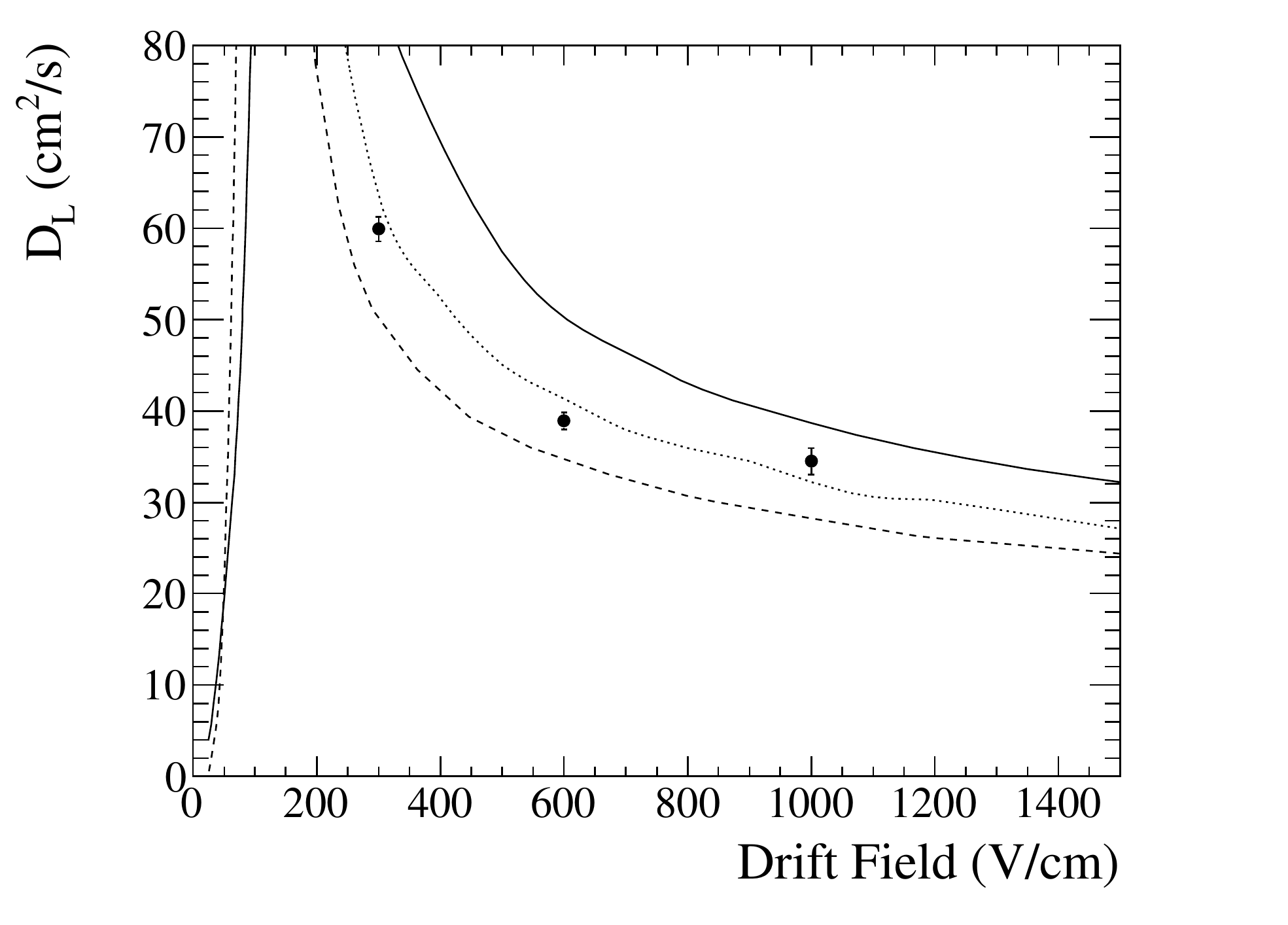}
\end{center}
\caption{Longitudinal diffusion coefficient $D_L$ as a function of drift field, for xenon gas at 10 bar pressure. The black points are the measured values. The dashed curve is the prediction for pure xenon from the Escada \textit{et al.} \protect\cite{Escada:2011cs}) simulation. The solid curve is the Magboltz \protect\cite{magboltz} prediction based on version 9.0.1 of the code, and include thermal motion of atoms. The dotted curve is based on a development version of the Magboltz code \protect\cite{biagi}. The error bars on the NEXT-DEMO data account for the main uncertainties (drift velocity and fit model, see text for details).}\label{fig:diffusionvsfield}
\end{figure}

Figure \ref{fig:diffusionvsfield} shows how our measured longitudinal diffusion coefficients $D_L$ for the 0.3, 0.6, and 1 kV/cm field configurations compare to three different models for pure xenon gas at 10 bar. The difference among predictions is much larger than the corresponding one for the drift velocity predictions (Fig.~\ref{fig:driftvelocityvsfield}), motivating the need for longitudinal diffusion measurements. Our results fall in between the current Magboltz \cite{magboltz} (version 9.0.1, atomic thermal motion included) and the Escada \textit{et al.} \cite{Escada:2011cs} predictions already mentioned in Sec.~\ref{subsec:ElectronTransportVelocity}. The third model refers to a development version of the Magboltz code meant to improve only the prediction of the longitudinal diffusion in pure xenon \cite{biagi}. The latter model describes our data best. Also, our longitudinal diffusion results are found to be in agreement with other NEXT R\&D results on diffusion, obtained with the NEXT-DBDM prototype \cite{nextdbdmpaper}.

The error bars on the NEXT-DEMO diffusion data assume two main contributions, summed in quadrature: a drift velocity uncertainty, plus a fit model uncertainty. Both contributions turn out to be of the same order. The drift velocity uncertainty is estimated as in Fig.~\ref{fig:driftvelocityvsfield}. The fit model uncertainty is estimated as the variation in the fit result with respect to nominal as the fit range in $z$ is extended to include also $\sigma_{t}<1.15\ \mu$s events.

 Finally, the best-fit constant term $\sigma_0$ in Eq.~\ref{eq:diffusionfit} shows little dependence with drift field, as expected, and values comprised between 0.77 and 0.81 $\mu$s for the three field configurations are obtained. The scale of both the $\sigma_0$ average values ($\simeq 0.8$ $\mu$s) and the width of the $\sigma_t$ distribution at small drift lengths ($\simeq 0.3$ $\mu$s, Fig.~\ref{fig:diffusionfit}) can be understood with simple arguments. For an alpha particle decaying perpendicularly to the drift direction and for negligible diffusion, the S2 signal can be approximated by a square pulse with a width given by the ionization electrons transit time through the EL region. For 10 bar xenon gas in a 10 kV/cm EL field, it takes about 2 $\mu$s to cross the 4.7 mm wide EL region, resulting in a lower bound on $\sigma_t$ of about $2/\sqrt{12}\simeq 0.6$ $\mu$s. On the other hand, the longest S2 pulses will be obtained from alpha particles decaying along the drift field direction. At 10 bar pressure, such decays generate an approximately constant flow of ionization electrons reaching the EL region of about 2 $\mu$s duration. The convolution of this square ionization input pulse with the EL response can be approximated by a 4 $\mu$s wide triangular S2 pulse, yielding $\sigma_t\simeq 4/(2\sqrt{6})\simeq 0.8$ $\mu$s, and therefore a 0.2 $\mu$s wide $\sigma_t$ distribution from alpha track orientation arguments alone. Another factor contributing to this non-zero ionization signal pulse time extent at zero drift length is the shaping of the electronics, with a time constant of about 0.2 $\mu$s. As diffusion becomes more important, the gaussian approximation for the S2 signal shape quickly becomes accurate, as shown in Fig.~\ref{fig:waveform}.

\section{Ionization yield, scintillation yield, and electron-ion recombination studies} \label{sec:Recombination}

In Sec.~\ref{sec:ElectronTransport}, we use observables in the time domain (namely, drift time and time extent of the ionization signal) to infer ionization electron transport properties. In this section, we use charge information to study the scintillation yields, the ionization yields, and the relationship between the two. As for the diffusion measurements in Sec.~\ref{subsec:ElectronTransportDiffusion}, for this we use the sample of bulk alpha candidate events described in Sec.~\ref{subsec:ProcessingSelectionGasAlphas}.

\subsection{Correlated fluctuations between ionization and scintillation} \label{subsec:RecombinationFluctuations}

The simultaneous measurement of ionization and scintillation allows for the study of correlated fluctuations among the two. Because of the small Fano factor of xenon gas (for $\rho \lesssim 0.5$ g/cm$^3$ densities), correlated fluctuations due to the partitioning of a fixed amount of deposited energy into these two pathways of energy loss are expected to be negligibly small in our case, at the $10^{-3}$ level. Once instrumental effects are corrected for, we can therefore interpret anti-correlated fluctuations as due to fluctuations in the electron-ion recombination process.

\begin{figure}[t!]
\begin{center}
\includegraphics[scale=.70]{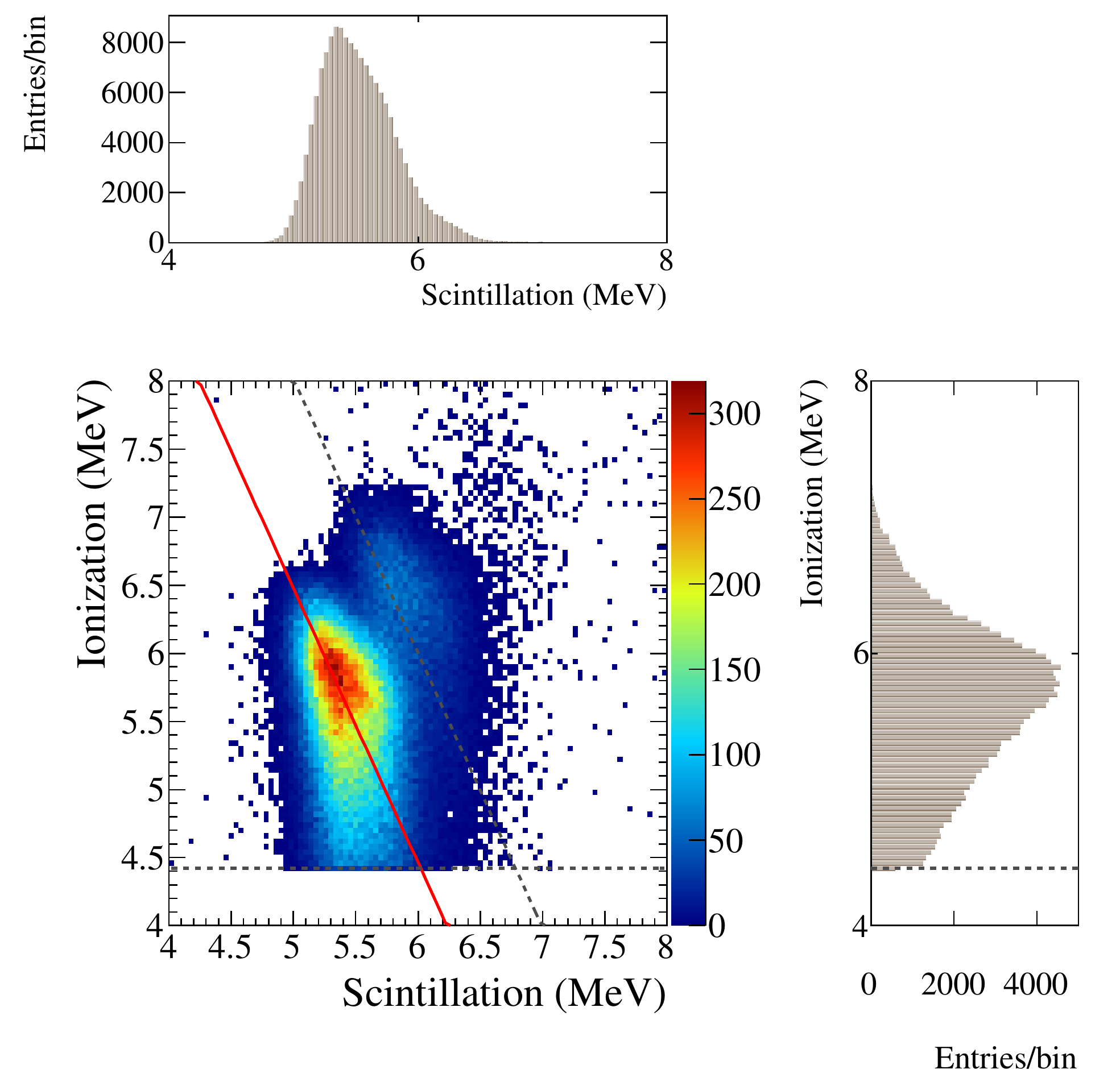}
\end{center}
\caption{Ionization (S2) signal yield as a function of scintillation (S1) signal yield for bulk alpha candidate events in the $E_{\textrm{drift}}=0.6$ kV/cm run (color histogram), illustrating the effect of correlated fluctuations between the two. The two yields have been corrected for the spatial effects described in Sec.~\protect\ref{subsec:ProcessingSelectionSpatialCorrections}, and also normalized to an absolute energy (MeV) scale. The events around $E_{\textrm{S1}}=5.5$ and $6.0$ MeV are mostly $^{222}$Rn and $^{218}$Po alpha decay events, respectively. The red curve is the $E=5.49$ MeV line, where the energy $E$ is defined as in Eq.~\protect\ref{eq:energy}. In order to extract the red line, only events to the left and bottom of the dashed diagonal ($E=6.0$~MeV) line are used. The horizontal line shows the analysis S2 threshold. Finally, the 1D projections on the ionization and scintillation axes are also shown.}
\label{fig:s2vss1}
\end{figure}

Figure \ref{fig:s2vss1} shows the ionization (S2) yield as a function of the scintillation (S1) yield, for bulk alpha candidate events collected in the 0.6 kV/cm field configuration. Both yields have been corrected for the spatial effects described in Sec.~\ref{subsec:ProcessingSelectionSpatialCorrections}. The projections along the two axes are also shown in Fig.~\ref{fig:s2vss1}. As hinted by the 2D histogram (and more clearly demonstrated below), the majority of the events correspond to alpha decays of $^{222}$Rn (5.49 MeV energy, see Fig.~\ref{fig:rn222decayscheme}), plus a $\sim$15\% fraction of alpha decays of $^{218}$Po (6.00 MeV energy). Large fluctuations are visible both in the ionization and scintillation response, preventing the identification of the $^{222}$Rn and $^{218}$Po alpha lines by using the S1 or S2 signals alone. Despite the fact that the number of photo-electrons detected for the ionization signal is about 50 times larger than the corresponding number for scintillation (see Figs.~\ref{fig:s1spatialcorr} and \ref{fig:s2spatialcorr}), fluctuations are larger for the ionization channel. However, the data also show a clear anti-correlation between the ionization and scintillation signals. For this reason, we adopt a similar procedure as the one followed in liquid xenon experiments to combine the two signals, in order to obtain a more accurate estimator of the deposited energy per event. Our procedure, described below, closely follows the one in \cite{Aprile:2007qd}.

In order to estimate the alpha energy, we proceed in the following, iterative, way. In a first step, we use the mean S1 and S2 values of the 2D histogram in Fig.~\ref{fig:s2vss1}, in PEs/PMT units for the S1 and S2 corrected yields at zero drift length ($\overline{S_1}$ and $\overline{S_2}$, respectively), to rescale the S1 and S2 response into absolute (MeV) energy units. In the following, we refer to these MeV-rescaled S1 and S2 yields as $E_{S1}$ and $E_{S2}$, respectively. At this stage, we assume that all events are 5.49 MeV alphas from $^{222}$Rn decay. In a second step, we fit the entire 2D $(E_{S1},E_{S2})$ histogram shown in Fig.~\ref{fig:s2vss1} to a 2D correlated gaussian distribution. The fit therefore includes six parameters: an overall normalization, mean and sigma of the S1 signal in MeV units, mean and sigma of the S2 signal in MeV units, and a dimensionless correlation coefficient $\rho_{S1,S2}$ between the two. From the fit results, we can estimate the S1-S2 correlation angle $\vartheta$ as:

\begin{equation}
\vartheta = \arctan{(\sigma_{S2}/\sigma_{S1})}
\label{eq:theta}
\end{equation}

\noindent where $\sigma_{S1}$ and $\sigma_{S2}$ are the gaussian fit sigma parameters along the two axes. In other words, $\vartheta$ is the angle with respect to the S1 axis of the major axis of the ellipse obtained from the 2D gaussian fit. The red solid line in Fig.~\ref{fig:s2vss1} is defined as the line having this slope $\vartheta$ with respect to the S1 axis, and passing through the mean $E_{S1}$ and $E_{S2}$ parameters returned by the fit. A better estimate of $\overline{S_1}$ and $\overline{S_2}$ (and therefore of $E_{S1}$ and $E_{S2}$, respectively, for each event) is also given by the fit, from the peak position of the 2D gaussian. In a third step, we repeat the 2D correlated gaussian fit to the $(E_{S1},E_{S2})$ histogram in Fig.~\ref{fig:s2vss1}, by removing the highest energy events. This is done to partially remove the biases introduced by $^{218}$Po alpha decays. In order to do so, we only consider events satisfying the $E<6$ MeV requirement, where the event energy estimator $E$ is obtained by performing a rotation in the $(E_{S1},E_{S2})$ plane according to the $\vartheta$ angle in Eq.~\ref{eq:theta}, as follows:

\begin{equation}
E = \frac{\sin\vartheta\cdot E_{S1}+\cos\vartheta\cdot E_{S2}}{\sin\vartheta +\cos\vartheta}
\label{eq:energy}
\end{equation} 

The $E<6$ MeV requirement, shown as the grey dashed line in Fig.~\ref{fig:s2vss1} that is parallel to the red solid line, was chosen as a compromise between maintaining a high efficiency for $^{222}$Rn alphas while rejecting as many $^{218}$Po events as possible (about half). The process is iterative in the sense that the 2D correlated gaussian fit to $E<6$ MeV events relies on estimates for the $\overline{S_1}$, $\overline{S_2}$ and $\vartheta$ parameters entering in Eq.~\ref{eq:energy} that are obtained from the previous iterative step. One or two additional iterations of this procedure have proven to be sufficient to obtain stable numerical results.

\begin{table}[t!b!]
\caption{Values of the parameters $\sigma_{S1}$, $\sigma_{S2}$ and $\rho_{S1,S2}$ for the 2D gaussian fit to bulk alpha candidate events in the $(E_{S1},E_{S2})$ plane, for the three drift field configurations studied. The value of the angle $\vartheta$ as defined in Eq.~\protect\ref{eq:theta} is also given.}\label{tab:2dgaussianfit}
\begin{center}
\begin{tabular}{ccccc}
\hline
Drift field & $\sigma_{S1}$    & $\sigma_{S2}$     & $\rho_{S1,S2}$     & $\vartheta$ \\
(kV/cm)     & (MeV)            & (MeV)            &                   & (rad) \\ \hline
0.3         & $0.303\pm 0.001$ & $0.521\pm 0.002$ & $-0.290\pm 0.004$ & $1.044\pm 0.001$    \\
0.6         & $0.247\pm 0.001$ & $0.497\pm 0.001$ & $-0.222\pm 0.004$ & $1.110\pm 0.001$    \\  
1.0         & $0.233\pm 0.001$ & $0.472\pm 0.002$ & $-0.213\pm 0.005$ & $1.112\pm 0.001$    \\ \hline
\end{tabular}
\end{center}
\end{table}

The parameters $\sigma_{S1}$, $\sigma_{S2}$ and $\rho_{S1,S2}$ of the 2D gaussian fit, together with the derived quantity $\vartheta$, are reported in Tab.~\ref{tab:2dgaussianfit} for the three drift field configurations studied. For all drift fields, the correlation coefficient is negative, and the resolution using the S1 signal alone is about a factor of 2 better than using the S2 signal alone. As expected, and as already observed in liquid xenon \cite{Conti:2003av, Aprile:2007qd}, the anti-correlation and the fluctuations are strongest for the lowest drift field configuration, corresponding to the strongest recombination scenario. We have also estimated the impact on the anti-correlation coefficient that could be attributed to instrumental effects and not to physics, namely from an inaccurate spatial correction to the S1 and S2 signals (see Sec.~\ref{subsec:ProcessingSelectionSpatialCorrections}). Taking as an example the 0.6 kV/cm drift field data, and by varying the S1 and S2 spatial correction coefficients within their error envelope, we obtain a $\pm 0.016$ uncertainty on the correlation coefficient. 

\begin{figure}[t!]
\begin{center}
\includegraphics[scale=.60]{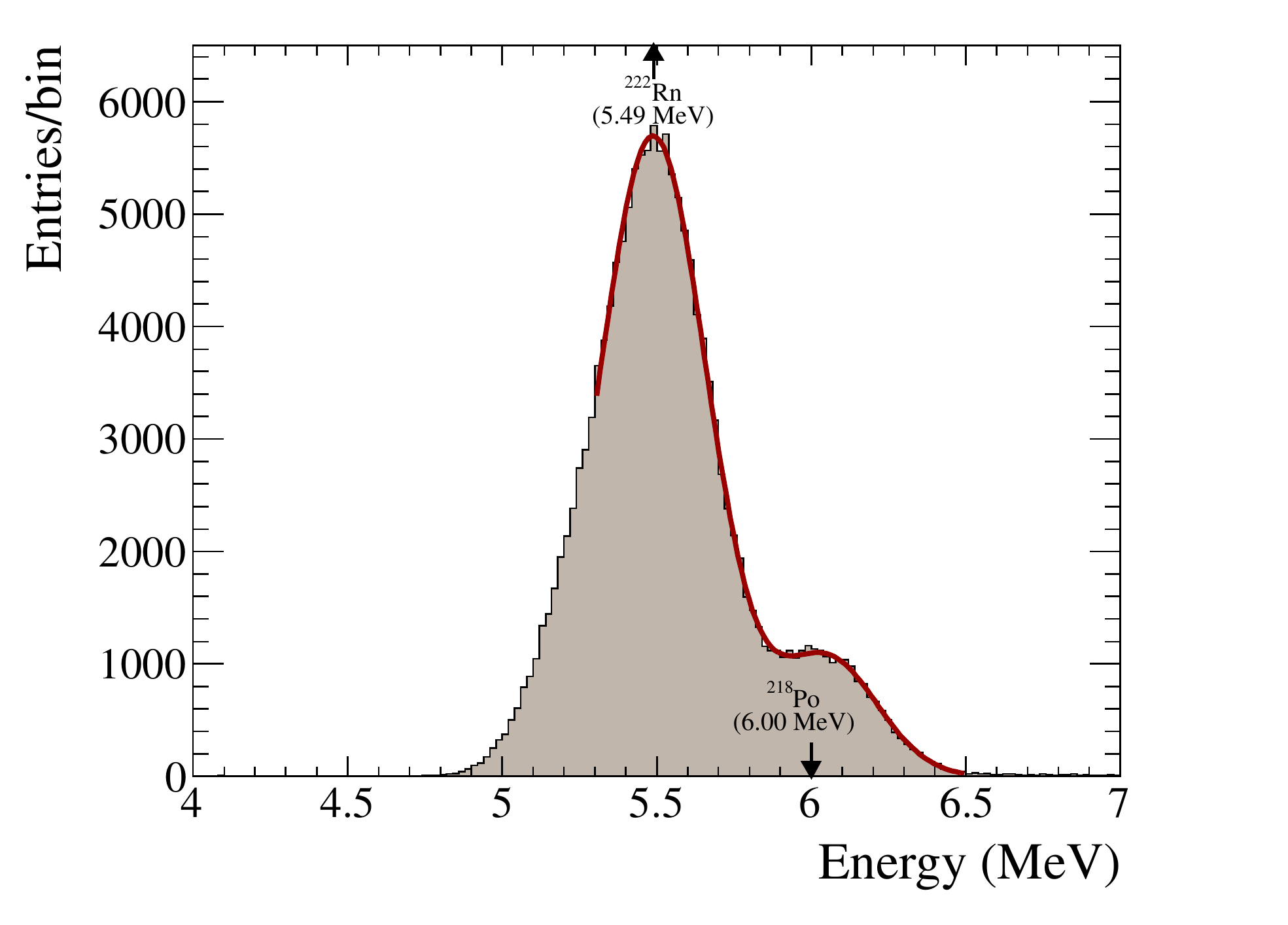}
\end{center}
\caption{Energy distribution for bulk alpha candidate events in the $E_{\textrm{drift}}=0.6$ kV/cm run, where the energy is defined according to Eq.~\protect\ref{eq:energy}. Two peaks, due to $^{222}$Rn and $^{218}$Po alpha decay events, are visible. The red curve is a fit to the data, assuming the sum of two gaussian functions to describe the two populations.}\label{fig:energy}
\end{figure}

The event energy $E$ for all bulk alpha candidate events collected during the 0.6 kV/cm drift field run is shown in Fig.~\ref{fig:energy}. Two peaks are clearly visible, corresponding to $^{222}$Rn and $^{218}$Po alpha decays, respectively. For energies above 5.3 MeV, the data are well described by the sum of two gaussian functions for these two populations. This fit is also shown in Fig.~\ref{fig:energy}. While the $^{222}$Rn peak position is essentially required by our procedure to be aligned with its expected value of 5.49 MeV, this is not the case for the $^{218}$Po peak. The fact that the latter appears near the expected 6.00 MeV value gives us confidence in our hypothesis that most events in our sample are $^{222}$Rn alpha decays, with about a 15\% fraction due to $^{218}$Po alpha decays. Both gaussian distributions in the fit of Fig.~\ref{fig:energy} are characterized by a FWHM energy resolution of about 8\%, slightly improving with increasing drift field. This energy resolution is significantly worse than the 1.75\% FWHM value obtained with the same NEXT-DEMO detector using a $^{22}$Na, 511 keV, gamma ray source \cite{nextdemopaper}. In addition to degraded energy resolution compared to \cite{nextdemopaper}, a non-gaussian energy response for energies below 5.3 MeV is also apparent. This low-$E$ tail stems from the large low-ionization tail that is visible in Fig.~\ref{fig:s2vss1}, which is only partially eliminated by the S2 charge threshold cut discussed in Sec.~\ref{subsec:ProcessingSelectionGasAlphas}. The cause for the degraded energy resolution and for the non-gaussian energy response is qualitatively understood via Monte Carlo simulations as due to the lack of tracking readout plane information in the analysis presented here (as opposed to \cite{nextdemopaper}), preventing the use of a tighter fiducialization of the events and/or of a $(x,y)$ position-dependent correction to the ionization and scintillation yields. Simulations also indicate that the impact on energy resolution of statistical fluctuations in the number of detected photons is in this case negligible compared to the non-uniformities in energy response discussed above. In the future, alpha decay data to be taken in NEXT-DEMO with a tracking readout plane are therefore expected to exhibit a much better energy resolution, a more gaussian energy response, and a stronger anti-correlation between ionization and scintillation. 

\begin{table}[t!b!]
\caption{Values of the parameters for the double gaussian fit to the energy spectra, for the three drift field configurations studied. The mean and sigma of the $^{222}$Rn ($^{218}$Po) gaussian are indicated with $\overline{E}_{\textrm{Rn}}$ and $\sigma_{\textrm{Rn}}$ ($\overline{E}_{\textrm{Po}}$ and $\sigma_{\textrm{Po}}$), respectively. The ratio of $^{218}$Po-to-$^{222}$Rn events is given by $N_{\textrm{Po}}/N_{\textrm{Rn}}$.}\label{tab:energyfit}
\begin{center}
\begin{tabular}{cccccc}
\hline
Drift field & $\overline{E}_{\textrm{Rn}}$ & $\sigma_{\textrm{Rn}}$ & $\overline{E}_{\textrm{Po}}$ & $\sigma_{\textrm{Po}}$ & $N_{\textrm{Po}}/N_{\textrm{Rn}}$ \\
(kV/cm) & (MeV) & (MeV) & (MeV) & (MeV) & \\ \hline
0.3         & $5.484\pm 0.001$ & $0.191\pm 0.002$ & $6.050\pm 0.004$ & $0.173\pm 0.003$    & $0.174\pm 0.004$ \\
0.6         & $5.488\pm 0.001$ & $0.178\pm 0.001$ & $6.049\pm 0.003$ & $0.165\pm 0.002$    & $0.171\pm 0.003$ \\  
1.0         & $5.492\pm 0.001$ & $0.171\pm 0.001$ & $6.048\pm 0.003$ & $0.163\pm 0.002$    & $0.161\pm 0.003$\\ \hline
\end{tabular}
\end{center}
\end{table}

As can be seen from the fit parameters listed in Tab.~\ref{tab:energyfit}, the energy spectra do not change in any appreciable manner as the drift field is increased from 0.3 to 1.0 kV/cm. As mentioned in Sec.~\ref{subsec:ExperimentalSetupConfiguration}, the depletion of $^{218}$Po decays in the gas with respect to $^{222}$Rn decays indicates that the majority of the positively charged $^{218}$Po ions accumulate on the negatively charged field grids and on the PTFE reflector panels. The $^{214}$Po decays are even further suppressed by our selection of bulk alpha candidate events, with essentially all $^{214}$Po ions located at the field grids and reflector panels.

\subsection{Field dependence of ionization and scintillation average yields} \label{subsec:RecombinationFieldDependence}

Apart from the study of correlated fluctuations discussed in Sec.~\ref{subsec:RecombinationFluctuations}, electron-ion recombination in xenon gas can also be investigated by measuring how the average ionization and scintillation yields vary with drift electric field intensity. For very intense fields, recombination is negligible. In this case, ionization charge collection is unaffected by the charge carriers in the gas, and the scintillation signal is entirely due to primary scintillation. As the field is lowered and recombination becomes important, part of the ionization is lost while the scintillation signal increases because of recombination luminescence. Different recombination models yield different trends in the way both signals vary with electric field intensity and with ionization density (see, for example, \cite{bolotnikov1999}). For the drift field and gas pressure conditions probed here, variations at the few \% level in both the average S1 and S2 yields produced by alpha particles are expected, see for example \cite{bolotnikov1999,kobayashi,Saito:2003dz}.

\begin{figure}[t!b!]
\begin{center}
\includegraphics[scale=.55]{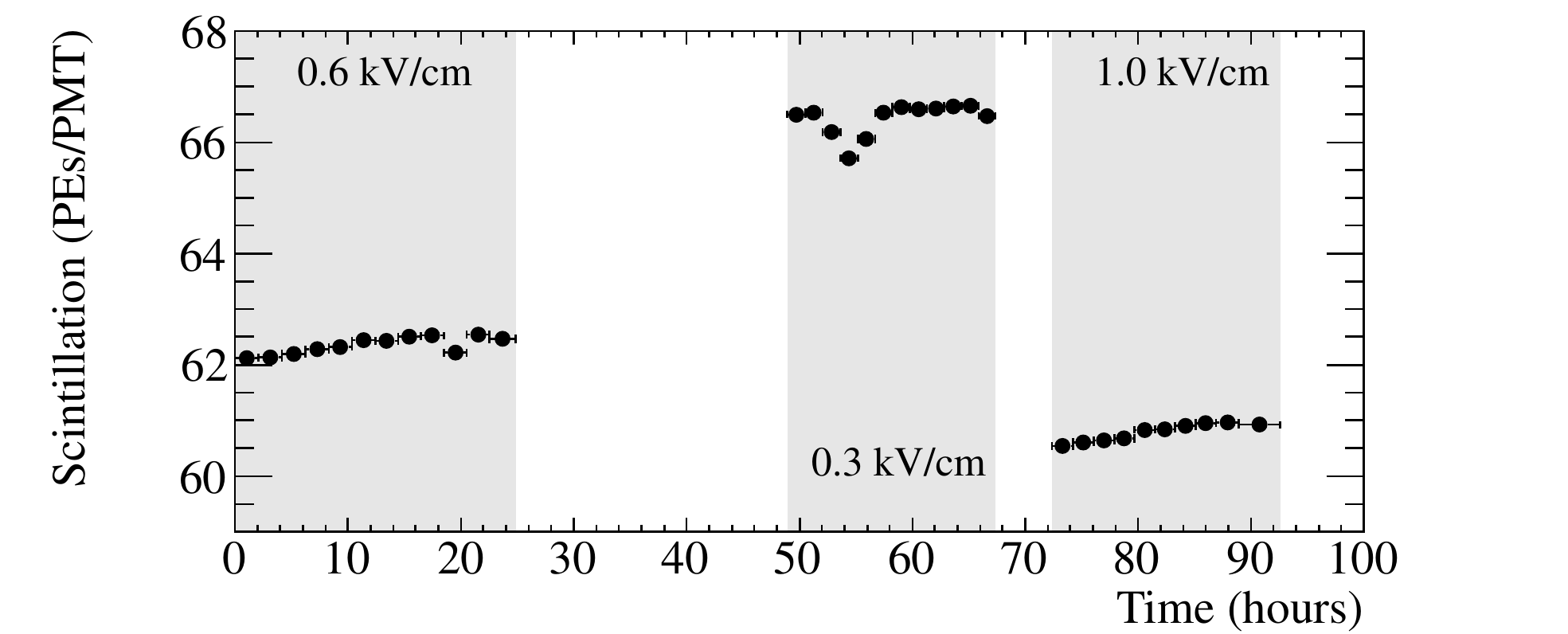}
\includegraphics[scale=.55]{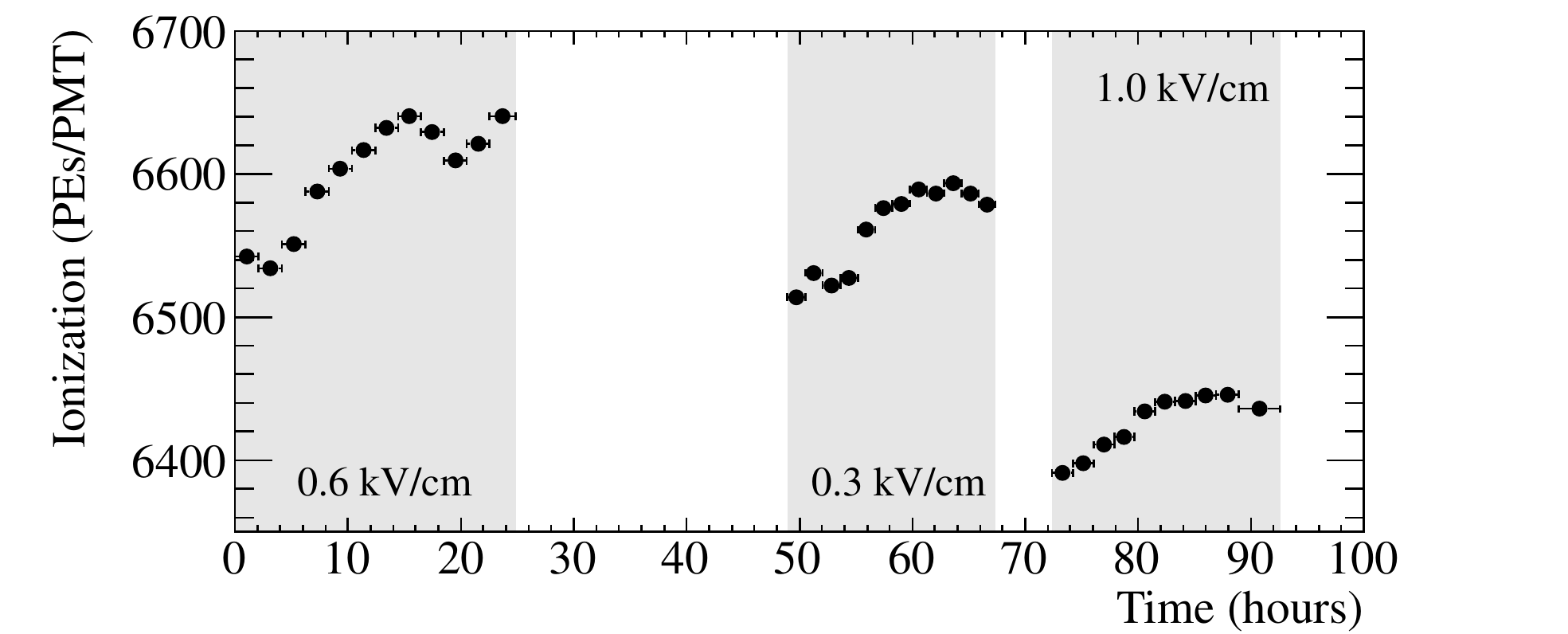}
\end{center}
\caption{Stability over time for the average scintillation (S1, upper panel) and ionization (S2, lower panel) signals of bulk alpha candidate events, for the three drift field configurations studied.}\label{fig:s1s2stability}
\end{figure}

In order to understand the instrumental uncertainties affecting our yield measurements, we show in Fig.~\ref{fig:s1s2stability} the stability over time of the average S1 and S2 yields (in PEs/PMT) for bulk alpha candidate events. The average yields are corrected for the spatial effects described in Sec.~\ref{subsec:ProcessingSelectionSpatialCorrections}. As a measure of stability for a given drift field configuration, we compute the average yields for two-hour periods during the run, and compute the relative standard deviation among them. With such measure, the average S1 (S2) yields are stable to within 0.4\% (0.6\%) or better, for all three drift field configurations probed. Also, Fig.~\ref{fig:s1s2stability} suggests a mild increase of the scintillation and ionization yields over time within a drift field configuration, somewhat more evident for the ionization case. The slope and magnitude of both the S1 and S2 trends are ultimately caused by the small pressure leak during the run (see Sec.~\ref{subsec:ElectronTransportVelocity}). 

In the scintillation case, we have seen that the yield greatly depends on the alpha decay $z$ position, with a factor of 2 yield variation in the $40<z\ \textrm{(mm)}<220$ range (see Fig.~\ref{fig:s1spatialcorr}). The small pressure decrease over time causes a small drift velocity increase over time (see Fig.~\ref{fig:driftvelocitystability}). This time variation in drift velocity is not accounted for in this analysis, ultimately causing a time-varying bias in the decay reconstructed $z$-positions and a time dependence in the $z$-corrected scintillation yields shown in Fig.~\ref{fig:s1s2stability}. 

In the ionization case, the yield dependence on $z$ position is much milder (see Fig.~\ref{fig:s2spatialcorr}). In this case, the main observable consequence of a time-varying pressure is a change in the EL gain of the detector. At the near EL threshold operating conditions of this run, a 0.5\% pressure decrease over one day is expected to cause approximately a 1\% increase in EL gain \cite{Oliveira:2011xk}, and therefore in the S2 yield. Unlike S1 yield and drift velocity, S2 yield and EL gain cannot be separately measured in our data. For this reason, and for the lack of archival pressure monitoring data during this run, we assume an additional 4\% uncertainty on the average S2 yield from one drift field configuration to another. This additional uncertainty is due to the effect that a xenon refill at the level of a 2\% gas pressure increase would have on the detector EL gain.

\begin{figure}[t!b!]
\begin{center}
\includegraphics[scale=.55]{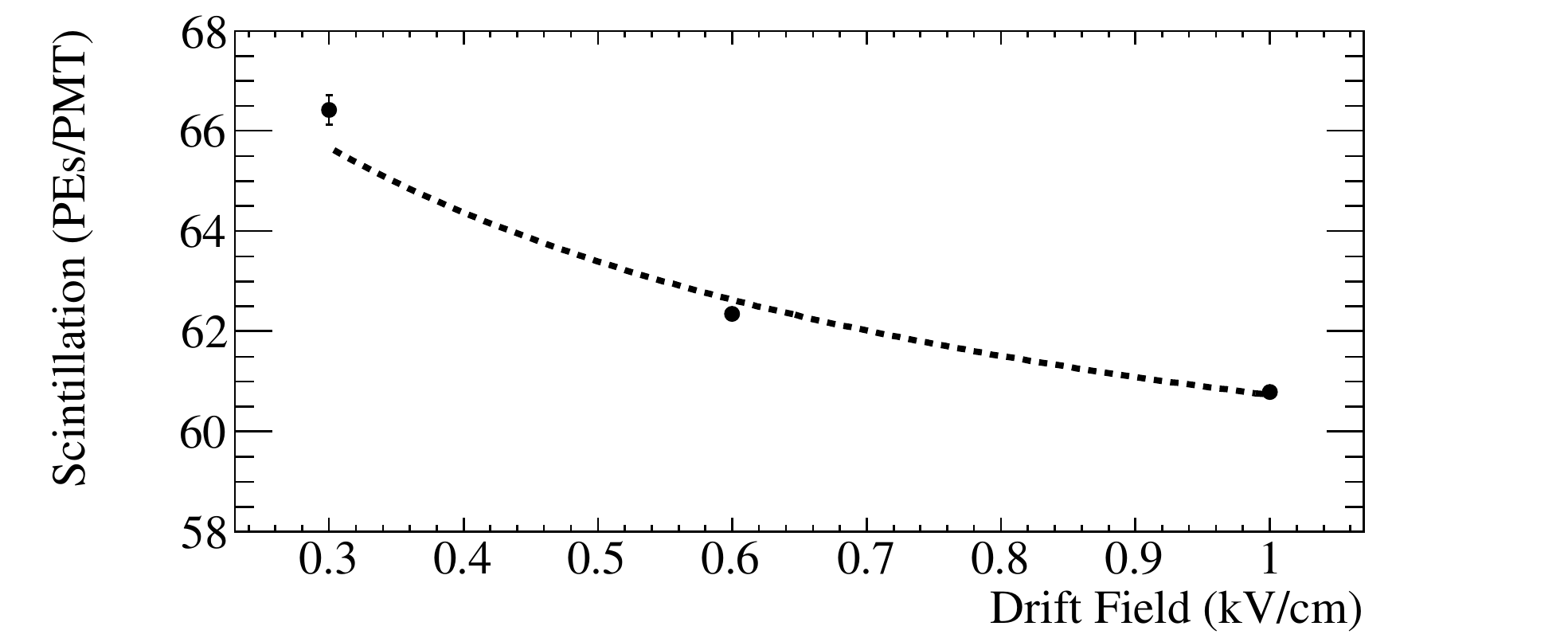}
\includegraphics[scale=.55]{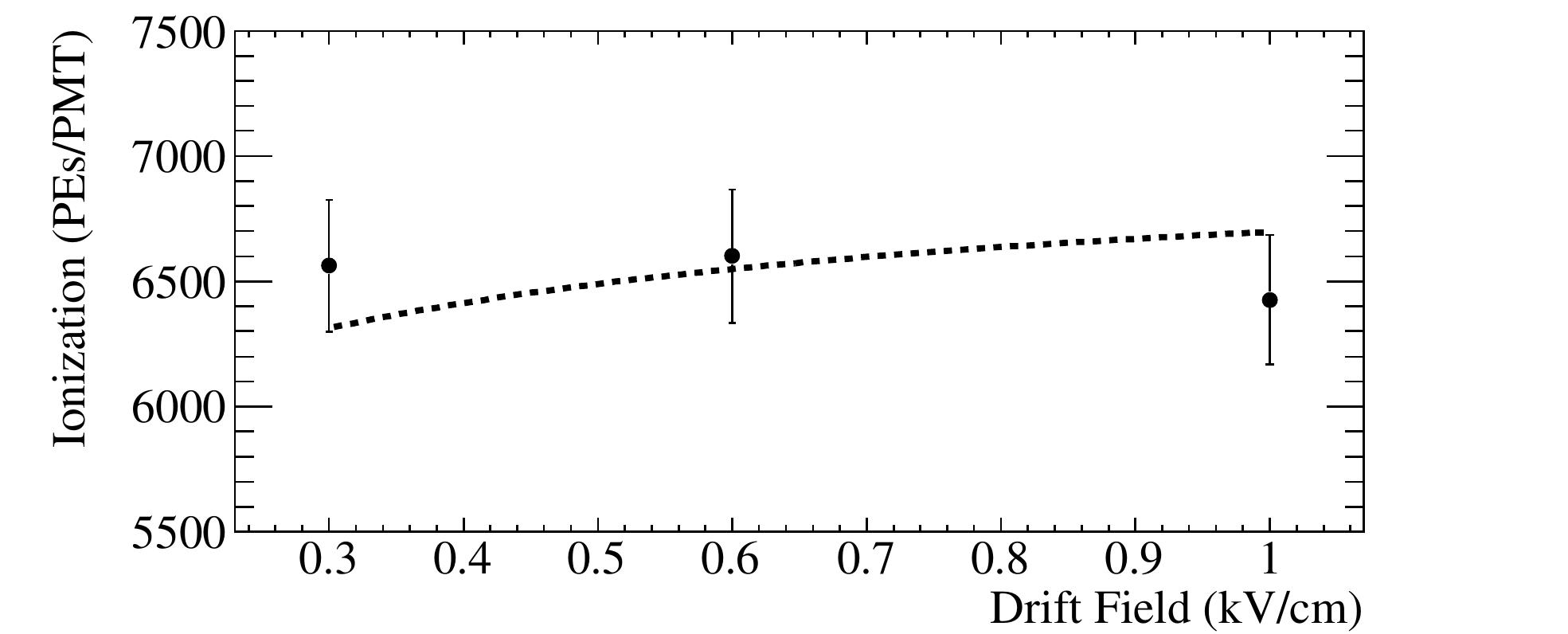}
\end{center}
\caption{Average scintillation (S1, upper panel) and ionization (S2, lower panel) yields as a function of drift field, for bulk alpha candidate events. The dashed lines are the best-fit results for the two-component recombination model of \cite{bolotnikov1999}.}\label{fig:means1s2vsfield}
\end{figure}

Figure~\ref{fig:means1s2vsfield} shows how the average scintillation and ionization yields for bulk alpha candidate events vary with drift field. The points in the figure are the mean of all measured yields over the two-hour time intervals at a given drift field in Fig.~\ref{fig:s1s2stability}. For the scintillation yields, the error bars are given by the standard deviation among these measured yields. For the ionization yield, the error bars are dominated by the 4\% EL gain uncertainty mentioned above. We observe that the average scintillation yield increases by $(9.3\pm 0.6)\%$ as the field is decreased from 1.0 to 0.3 kV/cm, in good agreement with \cite{kobayashi} and \cite{Saito:2003dz} for xenon gas in the same pressure conditions and also using alpha particles. On the other hand, and because of the large uncertainties, no clear trend can be appreciated in the average ionization yield as a function of drift field. 

Figure \ref{fig:means1s2vsfield} also shows a fit to the average scintillation and ionization yields as a function of drift field $E_{\textrm{drift}}$, for the two-component recombination model given in \cite{bolotnikov1999}. Following the notation in \cite{bolotnikov1999}, we express the average scintillation yield as:
\begin{equation}
L(E_{\textrm{drift}}) = L_{\textrm{ex}}+KQ_0(1-Q(E_{\textrm{drift}})/Q_0)
\label{eq:l}
\end{equation}
\noindent where $L_{\textrm{ex}}$ is the primary scintillation yield due to excitation of the xenon gas in the drift region, $Q_0$ is the ionization yield when recombination is negligible, and $K$ is an experiment-dependent constant specifying how the recombination luminescence signal $L(E_{\textrm{drift}})-L_{\textrm{ex}}$ scales with the amount of recombined charge $Q_0-Q(E_{\textrm{drift}})$. The average ionization yield can be written as:
\begin{equation}
Q(E_{\textrm{drift}}) = Q_1 + \frac{Q_2}{1+K_2/E_{\textrm{drift}}}
\label{eq:q}
\end{equation}
\noindent where $Q_1+Q_2=Q_0$, $Q_2/Q_0$ is the fraction of total charge that thermalizes close to their parent ions and may undergo initial (or geminate) recombination, $K_2$ is a recombination rate coefficient controlling how initial recombination varies with drift field, and $Q_1/Q_0$ is the fraction of charge that thermalizes far away from their parent ions, escapes initial recombination and may undergo volume (or columnar) recombination. Equation~\ref{eq:q} provides a good fit to 10 bar pressure xenon gas data for $E_{\textrm{drift}}\gtrsim 0.1$ kV/cm \cite{bolotnikov1999}, that is in the range of interest here, in which case all the fraction $Q_1/Q_0$ escapes volume recombination also, and is detected. For drift fields $\lesssim 0.1$ kV/cm, the first term in Eq.~\ref{eq:q} is also expected to vary with drift field strength.

In order to obtain stable fit results, we make two further assumptions based on external data. First, from \cite{bolotnikov1999}, we fix $Q_1/Q_0$ to 0.8 (and therefore $Q_2/Q_0=0.2$). Second, we assume:
\begin{equation}
K \equiv \frac{L(E_{\textrm{drift}})-L_{\textrm{ex}}}{Q_0-Q(E_{\textrm{drift}})} \simeq 1.6~\frac{L_{\textrm{ex}}}{Q_0}
\label{eq:k}
\end{equation}
While the first equality in Eq.~\ref{eq:k} follows directly from Eq.~\ref{eq:l}, the second approximate equality is obtained from external data on light and charge produced by alpha particles in 10 bar xenon gas over a very wide (0.03--1.34 kV/cm) drift field range \cite{Saito:2003dz}. In \cite{Saito:2003dz}, $L(E_{\textrm{drift}})-L_{\textrm{ex}}$ and $Q_0-Q(E_{\textrm{drift}})$ were measured to be  $0.97\cdot L_{\textrm{ex}}$ and $0.59\cdot Q_0$, respectively, at $E_{\textrm{drift}}=0.03$ kV/cm and by assuming $L_{\textrm{ex}}\simeq L(1.34~\textrm{kV/cm})$, $Q_0\simeq Q(1.34~\textrm{kV/cm})$. As a result, we obtain $K\simeq 1.6\cdot L_{\textrm{ex}}/Q_0$ in Eq.~\ref{eq:k}.

With these assumptions, we fit simultaneously scintillation and ionization data to Eqs.~\ref{eq:l} and \ref{eq:q}, respectively. The fit result is shown by the dashed curves in Fig.~\ref{fig:means1s2vsfield}. The goodness-of-fit is relatively poor, with a $\chi^2$ per degree of freedom of 12.3/3=4.1. The poor goodness-of-fit is driven by the non-observation of a ionization signal increase in the highest (1 kV/cm) drift field data set. The parameters obtained by the fit are: $L_{\textrm{ex}}=(56.4\pm 0.7)$ PEs/PMT, $Q_0=(7,040\pm 180)$ PEs/PMT and $K_2=(0.32\pm 0.07)$ kV/cm. 

Given the relatively low EL field used, $E_{\textrm{el}}\simeq 10.6$ kV/cm, we have also considered the possibility of loss of ionization electrons at the gate mesh because of non-perfect gate transparency. This possibility may particularly affect our highest drift field data sets, where the field ratio $E_{\textrm{el}}/E_{\textrm{drift}}$ is lowest, and may provide an explanation for the tension between data and fit results in Fig.~\ref{fig:means1s2vsfield}. According to the Buneman's formula \cite{buneman}, however, full gate transparency should be accomplished in our case (30 $\mu$m diameter wires at 0.25 mm separation) for $E_{\textrm{el}}/E_{\textrm{drift}}>2.2$. This condition is amply met in our case, therefore loss of ionization electrons at the gate appears unlikely.

\subsection{Quenching of primary scintillation light for alpha particles} \label{subsec:RecombinationScintillationQuenching}

In addition to allowing the study of electron-ion recombination, the measurement of the scintillation light yield can provide insight on the scintillation light quenching factor $q_{\mathrm s}(\alpha)$ for alpha particles, see Sec.~\ref{sec:Introduction}. In order to study $q_{\mathrm s}(\alpha)$, we compare the alpha-induced scintillation yield to the electron-induced one in the NEXT-DEMO detector. Both recombination and quenching may cause a different scintillation light response of alpha particles with respect to electrons. For sufficiently high electric fields, however, we have seen in Sec.~\ref{subsec:RecombinationFieldDependence} that recombination can be made small for alpha particles (and even more so for electrons), allowing to isolate the quenching effect on \textit{primary} scintillation light alone. In this case, the dimensionless scintillation light quenching factor for alpha particles can simply be written as:

\begin{equation}
q_{\mathrm s}(\alpha) \equiv \frac{\overline{S}_{\alpha}/E_{\alpha}}{\overline{S}_{e}/E_{e}} 
\label{eq:quenching}
\end{equation}

\noindent where $\overline{S}_{\alpha}$ ($\overline{S}_{e}$) is the average primary scintillation light yield for alpha particles (electrons) corresponding to a deposited energy in xenon gas of $E_{\alpha}$ ($E_{e}$).

\begin{figure}[t!b!]
\begin{center}
\includegraphics[scale=.37]{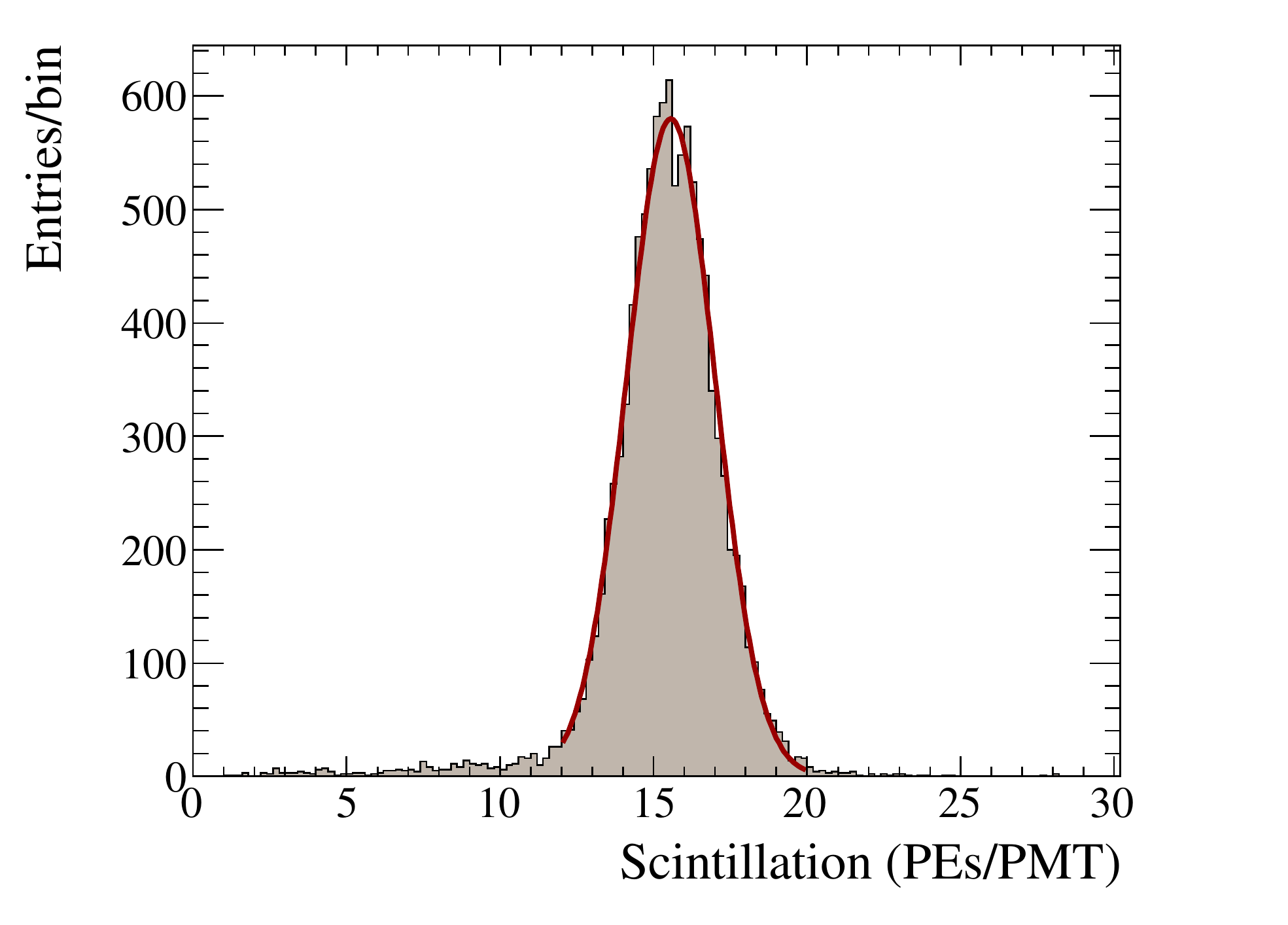}
\includegraphics[scale=.37]{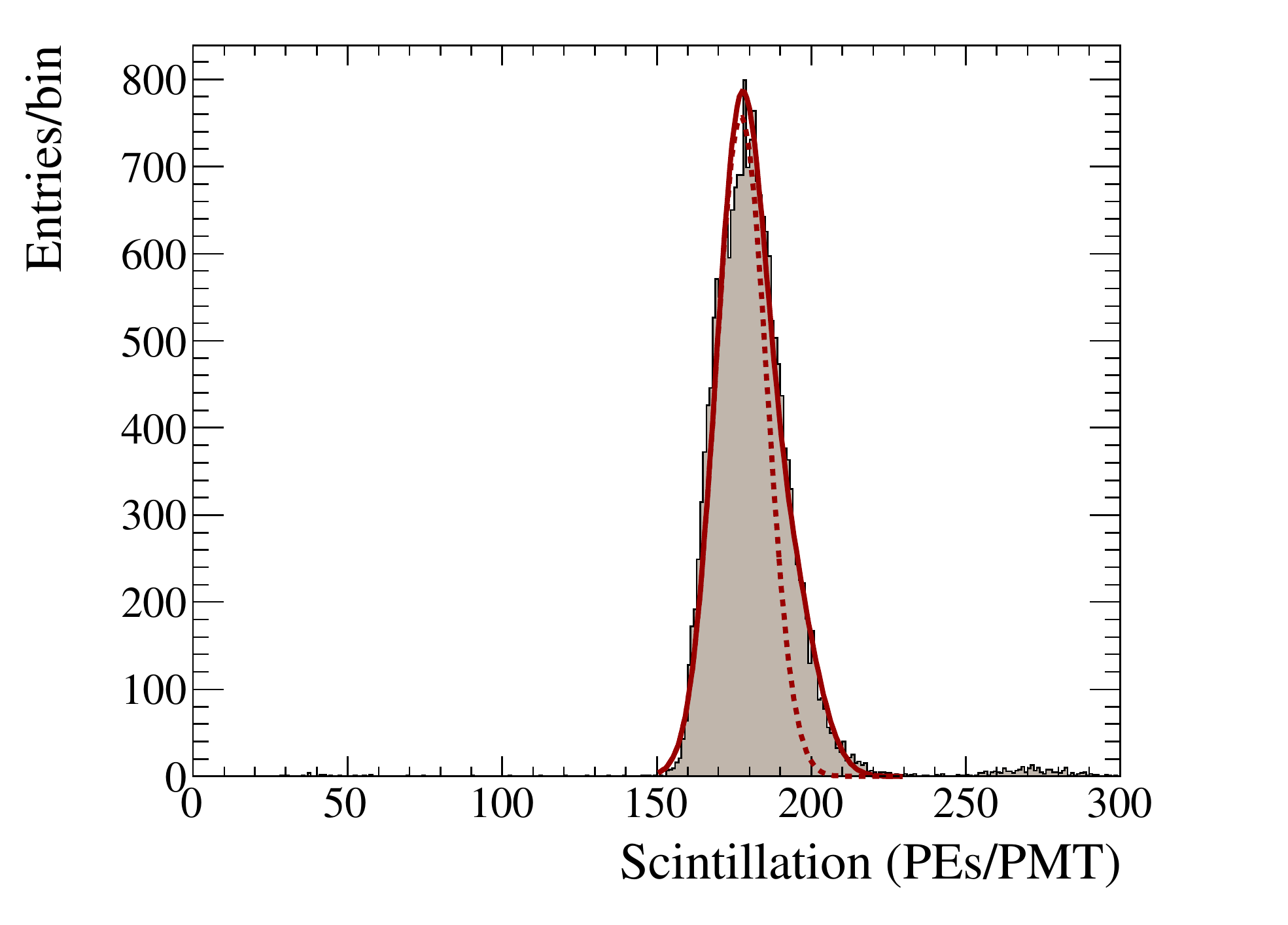}
\end{center}
\caption{Scintillation light yield for 0.511 MeV electrons (left panel) and 5.49--6.0 MeV alpha particles (right panel) in NEXT-DEMO. The red solid lines show gaussian fits to the data, as explained in the text. For the alpha case, the dashed red line shows the contribution from the $^{222}$Rn (5.49 MeV) gaussian alone.}\label{fig:s1quenching}
\end{figure}

Figure \ref{fig:s1quenching} shows the scintillation light response of NEXT-DEMO to both electrons and alpha particles. Both light responses are obtained with NEXT-DEMO reflector panels coated with TPB. For the electron spectrum, we consider the full-energy (photo-electric) events produced by 511 keV gammas from a $^{22}$Na source placed in a port near the TPC cathode, as reported in \cite{nextdemopaper}. Full-energy events are selected based on their ionization (S2) response. Events in the $260<z~\textrm{(mm)}<290$ drift distance range are selected. The region near the cathode was chosen for this comparison because that is where most 511 keV gammas interact, and also where the S1 yield exhibits essentially no dependence on the event $z$-position (see Fig.~\ref{fig:cathodeselection}). Any residual $z$-dependence is further suppressed in our sample by choosing this small, 30 mm wide, range in drift length. On the other hand, no cut on event radial (or $(x,y)$) position was imposed. The left panel in Fig.~\ref{fig:s1quenching} shows the S1 spectrum obtained for this sample, with no corrections applied. The spectrum is fitted to a gaussian, also shown in the figure. The best-fit values for the gaussian mean and sigma are $15.56\pm 0.01$ and $1.44\pm 0.01$ PEs/PMT, respectively. 

For the alpha spectrum to be used in the comparison, we consider bulk alpha candidate events. Events are selected as described in Sec.~\ref{subsec:ProcessingSelectionGasAlphas}, with the exception that the drift length range is now $260<z~\textrm{(mm)}<290$ (as for the electron sample) as opposed to $40<z~\textrm{(mm)}<220$. In order to minimize recombination effects, we use the highest drift field data set for this comparison, 1 kV/cm. The right panel in Fig.~\ref{fig:s1quenching} shows the S1 spectrum obtained for this sample, with no corrections applied. Since both $^{222}$Rn (5.49 MeV) and $^{218}$Po (6.0 MeV) alpha particles are present in the sample, the spectrum is fitted to the sum of two gaussian functions. Since the two peaks cannot be resolved, the $^{218}$Po peak position is constrained in the fit to scale according to deposited energy, with respect to the $^{222}$Rn peak. The other five parameters (two gaussian normalizations, two gaussian sigmas, and $^{222}$Rn gaussian mean) are kept free in the fit. Both the entire fit function as well as the $^{222}$Rn gaussian alone are also shown in the figure. The best-fit values for the $^{222}$Rn gaussian mean and sigma are $177.3\pm 0.2$ and $8.3\pm 0.1$ PEs/PMT, respectively. As quantified by the fit results in Sec.~\ref{subsec:RecombinationFieldDependence}, we estimate that a 93\% fraction of the luminescence signal at 1 kV/cm drift is due to primary scintillation. Therefore, we take $\overline{S}_{\alpha}=(164.9\pm 0.2)$ PEs/PMT in the following. The $^{218}$Po gaussian is found to have a width similar to the $^{222}$Rn one, as expected. Also, the ratio of $^{218}$Po to $^{222}$Rn events is $0.25\pm 0.02$. This ratio is somewhat higher than the value reported in Tab.~\ref{tab:energyfit}, suggesting that the concentration of $^{218}$Po events in the gas is not uniform, but higher near the cathode.

Therefore, from Eq.~\ref{eq:quenching}, and using $E_{e}=0.511$ MeV, $\overline{S}_{e}=15.56\pm 0.01$ PEs/PMT, $E_{\alpha}=5.49$ MeV, and $\overline{S}_{\alpha}=164.9\pm 0.2$ PEs/PMT, we obtain $q_{\mathrm s}(\alpha)=0.99$. We have also estimated the two main systematic uncertainties affecting our $q_{\mathrm s}(\alpha)$ measurement. First, we account for the fact that the NEXT-DEMO configuration was slightly different for the $^{22}$Na and alpha runs. In the $^{22}$Na case, a PMT-based tracking readout plane was installed near the anode. In the alpha case, a TPB-coated surface was present. According to Monte Carlo simulations, the latter (and more reflective) configuration should produce about 8\% more scintillation light detected at the energy readout plane, for energy depositions occurring near the cathode. Conservatively, we take this difference as a systematic uncertainty. Second, we assign an uncertainty associated with the different radial distribution of electron and alpha events near the cathode, causing the energy readout plane response to scintillation light to be slightly different in the two cases. By studying how the scintillation light yield for electron events and for simulated point-like energy depositions vary in different radial shells, we estimate a 8\% uncertainty associated to the different radial distribution of events, also. Overall, our result for the scintillation light quenching factor for alpha particles in xenon gas at 10 bar pressure is therefore $q_{\mathrm s}(\alpha)=0.99\pm 0.12$. In other words, given that $q_{\mathrm s}(\alpha)$ is consistent with unity, no scintillation light quenching is observed.
\section{Conclusions} \label{sec:Conclusions}

During the past few years, the NEXT Collaboration has conducted an R\&D program geared toward the construction of a high-pressure xenon gas time projection chamber (TPC) at the 100 kg scale (NEXT-100), to perform a sensitive search for neutrinoless double beta decay in $^{136}$Xe. Several time projection chamber prototypes have been built in this context, to prove the technological viability of the proposed NEXT-100 detector concept and to reliably quantify its projected performance. NEXT-DEMO is the largest of such prototypes, and has proven to be an ideal tool to also study the detection properties of high-pressure xenon gas in a broader context.

In this paper, we have studied both the ionization and the scintillation response of 10 bar pressure xenon gas to alpha particles, operating the NEXT-DEMO time projection chamber at 0.3--1.0 kV/cm drift field strengths. Alpha particles from the decay of the $^{222}$Rn, $^{218}$Po and $^{214}$Po isotopes are introduced in the detector, by allowing the xenon gas to flow through a $^{226}$Ra radioactive source. Two clean samples of alpha candidate events are used, one from decays in the xenon gas and one from decays from the TPC cathode.

Alpha particles are used to study the transport properties of ionization electrons, the process of electron-ion recombination, and scintillation light quenching in xenon gas. Concerning electron transport, an accurate monitoring of electron drift velocity was obtained using alpha decays from the cathode surface, confirming model predictions for pure xenon to the few percent level. The longitudinal diffusion of ionization electrons as they drift has also been accurately measured, by exploiting the point-like nature of alpha-induced energy depositions in the gas. Longitudinal diffusion is less well understood than electron drift velocity, and transport models can differ by as much as a factor of two in their calculation of the longitudinal diffusion coefficient \cite{lxcat}, motivating the need for accurate longitudinal diffusion measurements such as ours. Our result has a direct impact on the NEXT-100 physics potential, since the signal-to-background discrimination achievable in the detector by using event topology is affected by electron diffusion.

The process of electron-ion recombination can be best understood by the simultaneous measurement of both the scintillation and the ionization response. When recombination is negligible, the observed scintillation light is entirely primary, and no ionization charge losses (apart from attachment) occur. As electron-ion recombination becomes non-negligible, part of the ionization signal is lost to the benefit of a stronger scintillation response. The amount of recombination is expected to increase with decreasing electric field and with increasing linear energy transfer, the latter being defined as the quotient between the particle energy deposition per unit path length and the density of the detection medium. Despite the relatively low density of 10 bar pressure xenon gas (that is, low density relative to liquid xenon), the use of highly ionizing alpha particles allows us to clearly observe recombination in xenon gas in two different ways. On the one hand, we observe a reduction of the average scintillation response of xenon gas as the field is increased, as already observed several times in the past \cite{pushkin, bolotnikov1999, kobayashi, Saito:2003dz}. The scintillation light response to 5.49 MeV alpha particles from $^{222}$Rn decreases by about 10\% by increasing the drift field from 0.3 to 1.0 kV/cm. On the other hand, we observe large event-by-event correlated fluctuations between the ionization and scintillation response. We measure negative correlation coefficients, and a stronger anti-correlation for lower drift fields, as expected from recombination. We exploit this anti-correlation to build a better energy estimator that combines both the ionization and scintillation response. While similar findings and strategies have been reported in the past in liquid xenon \cite{Conti:2003av, Aprile:2007qd}, to our knowledge this is the first time that event-by-event correlated fluctuations between ionization and scintillation are observed in xenon gas. 

The scintillation light response of xenon gas to alpha particles may not only be affected by recombination, but also by quenching effects. We have studied quenching by exploring how the primary scintillation light yields of alpha particles and electrons \cite{nextdemopaper} compare. We find no evidence for quenching of the primary scintillation light produced by alpha particles in xenon gas at 10 bar pressure, within the $\sim$10\% accuracy of our measurement.

Finally, in addition to gaining a better understanding of the detection properties of xenon gas, the analysis techniques described here have a direct application toward both low-level monitoring and background measurement tools for NEXT-100. Concerning monitoring of key operational parameters, we have discussed how alpha particles can be used to track time variations at the percent level or better in the relative electroluminescent gain and, especially, in the electron drift velocity. Also, radon gas diffusing in the detector is a potential background to the neutrinoless double beta decay search, see for example \cite{:2012haa}. The work presented here is a first step toward establishing techniques to directly measure the concentration of radon gas in the detector.


\acknowledgments
This work was supported by the following agencies and institutions: the Spanish Ministerio de Econom\'ia y Competitividad under grants CONSOLIDER-Ingenio 2010 CSD2008-0037 (CUP) and FPA2009-13697-C04-04; the Portuguese FCT and FEDER through the program COMPETE, projects PTDC/FIS/103860/2008 and PTDC/FIS/112272/2009; and the Director, Office of Science, Office of Basic Energy Sciences, of the U.S. Department of Energy under Contract No. DE-AC02-05CH11231. J.~Renner (LBNL) acknowledges the support of a US DOE NNSA Stewardship Science Graduate Fellowship under contract no. DE-FC52-08NA28752.

\end{document}